\documentclass[%
preprint,
bibnotes,
amsmath,amssymb,
  aps,
prmaterials,
 prb,   %uncomment this for superscript citations
]{revtex4-1}

\def\A{\text{\AA}}
\def\kb{k_\text{B}}

\usepackage{graphicx}% Include figure files
\usepackage{dcolumn}% Align table columns on decimal point
\usepackage{ragged2e}

\usepackage{bm}% bold math
\usepackage{physics}
\usepackage{color}
\usepackage{subcaption}

\begin{document}

\title{Uncertainty quantification in first-principles predictions of phonon properties and lattice thermal conductivity}

\author{Holden L. Parks}
\affiliation{Department of Mechanical Engineering, Carnegie Mellon University, Pittsburgh, Pennsylvania 15213, USA}

\author{Hyun-Young Kim}%
\affiliation{Department of Mechanical Engineering, Carnegie Mellon University, Pittsburgh, Pennsylvania 15213, USA}

\author{Venkatasubramanian Viswanathan}%
% \email{venkvis@cmu.edu}
\author{Alan J. H. McGaughey}%
\email{mcgaughey@cmu.edu}
\affiliation{Department of Mechanical Engineering, Carnegie Mellon University, Pittsburgh, Pennsylvania 15213, USA}

\date{\today}

\begin{abstract}
We present a framework for quantifying the uncertainty that results from the choice of exchange-correlation (XC) functional in predictions of phonon properties and thermal conductivity that use density functional theory (DFT) to calculate the atomic force constants. The energy ensemble capabilities of the BEEF-vdW XC functional are first applied to determine an ensemble of interatomic force constants, which are then used as inputs to lattice dynamics calculations and a solution of the Boltzmann transport equation. The framework is applied to isotopically-pure silicon. We find that the uncertainty estimates bound property predictions (e.g., phonon dispersions, specific heat, thermal conductivity) from other XC functionals and experiments. We distinguish between properties that are correlated with the predicted thermal conductivity [e.g., the transverse acoustic branch sound speed ($R^2=0.89$) and average Gr{\"u}neisen parameter ($R^2=0.85$)] and those that are not [e.g., longitudinal acoustic branch sound speed ($R^2=0.23$) and specific heat ($R^2=0.00$)]. We find that differences in ensemble predictions of thermal conductivity are correlated with the behavior of phonons with mean free paths between $100$ and $300$ nm. The framework systematically accounts for XC uncertainty in phonon calculations and should be used whenever it is suspected that the choice of XC functional is influencing physical interpretations.
\end{abstract}

\maketitle

\section{\label{sec:intro}Introduction}

\textit{Ab initio} predictions of the lattice thermal conductivity of crystalline materials have become increasingly widespread due to their accuracy when compared to experimental measurements.\cite{broido, alan_jap, lindsay3} The prediction framework relies on density functional theory (DFT) to determine interatomic force constants, which quantify the change in potential energy of a material when its atoms are displaced from their equilibrium positions. The force constants are then used to predict harmonic phonon properties such as dispersion relations, group velocities, specific heat, and, by solving the phonon Boltzmann transport equation (BTE), anharmonic properties like scattering rates. \cite{alan_jap} These properties are then used to predict the thermal conductivity. This approach has been successful in studies of both low and high thermal conductivity materials, \cite{lindsay, shiga, ward,feng_2015, li, broido} and the predictions can be performed using one of several open-source packages. \cite{alamode, almabte, phonopy, shengbte} Models have grown more sophisticated in recent years to capture increasingly complex phenomena and resolve discrepancies between predictions and experimental values. \cite{Lindsay2019} For example, while it is common practice to consider only three-phonon scattering processes in solving the BTE, recent work has shown that four-phonon scattering is strong enough to reduce the predicted thermal conductivity in a range of materials. \cite{yang,feng,feng_ba, tian, ravichandran} There has also been progress in the treatment of finite temperature phases, \cite{finite_temp1, finite_temp2, finite_temp3, finite_temp4} compositionally-disordered materials, \cite{alloy, alloy2, alloy3, alloy4} and defects. \cite{defect,dislocation,guo}

While the continued improvement of models is important, the quality of the predictions also depends on computational parameters. \cite{alan_jap, ankit, xie} Obtaining converged phonon lifetimes, for example, requires a sufficiently-large cutoff radius for anharmonic interactions and a sufficiently-dense phonon wave vector grid. Similarly, the quality of the required force constants depends on the quality of the DFT calculation, which in turn is affected by a variety of factors. Some of these factors, such as the electronic wave vector grid and the plane wave energy cutoff (if a plane wave basis is used), require convergence testing. The exchange-correlation (XC) functional, conversely, is a critical component of all DFT calculations that must simply be chosen, sometimes with no \textit{a priori} knowledge of the most suitable selection.

The impact of XC functional choice, which we call the XC uncertainty, can be examined by changing the XC functional in otherwise identical thermal conductivity calculations. Jain and McGaughey calculated the thermal conductivity of silicon at a temperature of 300 K using five different XC functionals. \cite{ankit} They showed that the predictions could be as low as 127 W/m-K or as high as 172 W/m-K, in comparison with the experimental value of 153 W/m-K. \cite{si_expt} They argued that the variation was a result of differences in predictions of group velocities, three-phonon scattering phase space, and anharmonic effects. Taheri et al. calculated the thermal conductivity of graphene using three XC functionals and found the predictions to range from 5400 to 8700 W/m-K at a temperature of 300 K. \cite{taheri} Qin et al. also performed calculations on graphene at the same temperature using eight different XC functionals and reported a thermal conductivity range of 1900 to 4400 W/m-K. \cite{qin} In contrast to the silicon study by Jain and McGaughey, both Taheri et al$.$ and Qin et al$.$ found that all XC functionals tested agreed in predicting harmonic properties such as group velocity and three-phonon phase space, implying that the anharmonic properties are responsible for the large range of predicted thermal conductivities in graphene. These three previous studies demonstrate the impact of XC functional choice in predicting thermal conductivity, but the results are not conclusive indicators of XC uncertainty because the XC functionals tested were somewhat arbitrary. Additionally, this approach is computationally inefficient because nearly identical calculations must be performed for each XC functional choice.  Given the large number of possible functionals, even within the generalized gradient approximation (GGA) space, this brute force approach becomes computationally infeasible.

The Bayesian error estimation functional with van der Waals (BEEF-vdW) correlation is an XC functional that can systematically estimate XC uncertainty in DFT energies. \cite{beef} It possesses built-in uncertainty estimation capabilities in the form of an ensemble of GGA XC functionals that are calibrated to reproduce the discrepancies observed between experimental measurements and DFT predictions. The estimate obtained from the BEEF-vdW ensemble is computationally efficient because results for thousands of XC functionals are obtained non-self-consistently through a single self-consistent calculation. BEEF-vdW has been applied to quantify XC uncertainty in predictions of molecular vibrational frequencies, \cite{beef_molecules}  magnetic ground states, \cite{houchins}, intercalation energies, \cite{Pande2018} heterogeneous catalysis, \cite{medford, krishnamurthy2019, christensen} electrocatalysis, \cite{deshpande, krishnamurthy, christensen2} mechanical properties of solid electrolytes, \cite{ahmad} and thermodynamic properties. \cite{guan} Such uncertainty estimates are useful in machine learning-based materials design applications. For example, knowing the uncertainty associated with a DFT calculation can improve the robustness of workflows that rely on \textit{ab initio} calculations to screen materials.  \cite{Ramprasad2017,Ulissi2017}

In this work, we present a framework to estimate the XC uncertainty in predictions of phonon properties, specific heat, and thermal conductivity. The calculation details are presented in Sec.~\ref{sec:methods}. While we include only three-phonon scattering to save on computational cost, the framework can be easily extended to account for four-phonon and other scattering mechanisms. The framework is then applied in Sec.~\ref{sec:silicon} to isotopically-pure silicon, which is chosen due to its popularity as a benchmark for thermal conductivity predictions. \cite{alan_jap, ankit, esfarjani, broido} For comparison to the BEEF-vdW ensemble results, predictions of phonon properties, specific heat, and thermal conductivity using the LDA, \cite{lda} PBE, \cite{pbe} PBEsol, \cite{pbesol} and optPBE-vdW \cite{optpbe} XC functionals are also presented. We find that the BEEF-vdW ensemble accurately describes the variation of the self-consistent DFT predictions, with most predicted quantities bounded to within two ensemble standard deviations of the BEEF-vdW predictions. Based on analysis of the BEEF-vdW ensemble, we find that the best predictors of silicon thermal conductivity are the transverse acoustic phonon sound speed and $\mathbf{X}$-point frequency and the average Gr{\"u}neisen parameter. We also demonstrate the sensitivity of the thermal conductivity prediction to contributions of phonons with mean free paths in the range of $100$ to 300 nm.

\section{\label{sec:methods}Methods}

\subsection{\label{sec:beef}Bayesian error estimation}

BEEF-vdW is a semi-empirical XC functional that provides a way to systematically estimate the XC uncertainty in a DFT calculation. Its model space for the exchange-correlation energy $E_{\text{XC}}$ is given by \cite{beef}

\begin{equation}\label{eqn: beef model space}
E_{\text{XC}} = \sum_{m=0}^{29}\left (a_mE_m^{\text{GGA-x}}\right)+\alpha_cE^{\text{LDA-c}}+(1-\alpha_c)E^{\text{PBE-c}}+E^{\text{nl-c}}.
\end{equation}

\noindent Here, $E^{\text{LDA-c}}$, $E^{\text{PBE-c}}$, and $E^\text{nl-c}$ are the correlation contributions from the local Perdew-Wang LDA correlation, \cite{lda} the semi-local Perdew-Burke-Ernzerhof (PBE) correlation, \cite{pbe}  and the vdW-DF2 nonlocal correlation. \cite{df2} $E_m^{\text{GGA-x}}$ is the contribution to the exchange energy and is given by

\begin{equation}
E_m^{\text{GGA-x}} = \int \epsilon_\text{x}^\text{UEG}[n(\mathbf{r})] B_m \{t[n(\mathbf{r}), \nabla n(\mathbf{r})]\}d\mathbf{r},
\end{equation}

\noindent where $n(\mathbf{r})$ and $\nabla n(\mathbf{r})$ are the electron density and its gradient, $t$ is a function taking $n(\mathbf{r})$ and $\nabla n(\mathbf{r})$ as its inputs, $\epsilon_\text{x}^\text{UEG}$ is the exchange energy density of the uniform electron gas, and $B_m$ is the $m$th Legendre polynomial.

Wellendorff et al. fit the parameters $a_m$ ($0\leq m\leq 29$) and $\alpha_c$ to experimental training data to determine the optimal, or ``best-fit,'' BEEF-vdW XC functional. \cite{beef} The training data included six different data sets consisting of molecular formation and reaction energies, molecular reaction barriers, non-covalent interactions, solid-state properties such as cohesive energies and lattice constants, and chemisorption energies on solid surfaces. The training data did not include vibrational frequencies or thermal conductivities, so that our work will also serve as a test of the transferability of BEEF-vdW for predicting these properties.

BEEF-vdW provides a systematic and computationally-efficient approach to estimate the XC uncertainty in a DFT energy calculation through an ensemble of XC functionals. The electron density is first obtained through a self-consistent DFT calculation using the best-fit functional. An ensemble of XC functionals, each of which has its own set of $a_m$ and $\alpha_c$, are then applied to that electron density to yield an ensemble of non-self-consistent XC energies using Eq.~\eqref{eqn: beef model space}. Wellendorff et al. generated the ensemble of XC functionals using the following method. For every batch of data in the training set:

\begin{enumerate}

\item Use the best-fit XC functional to predict the values of interest and compare them with the experimental values. Call the sample standard deviation of these differences $s_1$.

\item Use the XC functional ensemble to predict the values of interest. Call the sample standard deviation of the ensemble $s_2$.

\end{enumerate} 

\noindent By tuning the distributions of $a_m$ and $\alpha_c$, Wellendorff et al. set $s_1$ and $s_2$ to be approximately equal, so that the spread in the ensemble recreates the differences between the experimental data and the best-fit BEEF-vdW predictions.
 
While the BEEF-vdW XC functional ensemble was generated to recreate differences between experimental and DFT data, there is no guarantee of its suitability for predicting properties not considered in the original training data set. The results of subsequent studies, \cite{houchins, ahmad, beef_molecules, medford, krishnamurthy2019, christensen, deshpande, krishnamurthy, christensen2, guan} however, demonstrated that the ensemble can reliably describe the XC uncertainty in self-consistent DFT predictions of a wide range of systems and material properties. In other words, the ensemble is transferable, in the sense that the variation in most self-consistent predictions is bounded in an interval of a few ensemble standard deviations. This result likely emerges because the ensemble exchange enhancement factors are similar to other common GGA-level functionals for reduced density gradient ($s=|\nabla n|/2k_\text{F}n$, where $k_\text{F}$ is the Fermi wave vector for a uniform electron gas) values between 0 and 2, \cite{beef} a range that describes most important interactions in chemical and solid-state systems. \cite{beef, hammer, csonka}

\subsection{\label{sec:phonons}Phonon properties and lattice thermal conductivity}

\subsubsection{\label{sec:k equation}Lattice thermal conductivity}

The phonon contribution to the thermal conductivity of a crystalline solid, i.e., the lattice thermal conductivity in direction $l$, $k_l$, can be obtained by solving the BTE in combination with the Fourier law and is given by \cite{alan_jap}

\begin{equation}\label{eqn: lattice k}
k_l = \sum_{\mathbf{q}, \nu} c (\mathbf{q}, \nu)v_{g,l}^2(\mathbf{q}, \nu)\tau_l(\mathbf{q}, \nu) . 
\end{equation}

\noindent Here, $\mathbf{q}$ and $\nu$ are the phonon wave vector and polarization, $c$ is the volumetric specific heat, and $v_{g,l}$ and $\tau_l$ are the group velocity and lifetime in the $l$ direction. The specific heats and group velocities are calculated using harmonic lattice dynamics, while the lifetimes require a combination of anharmonic lattice dynamics, perturbation theory, and the BTE. We only briefly discuss these calculations here as they have been described in detail elsewhere. \cite{McGaughey2014, alan_jap}

\subsubsection{\label{sec:hld}Harmonic lattice dynamics}

By assuming the phonon modes to be non-interacting plane waves, the frequencies $\omega$ and eigenvectors $\mathbf{e}$ associated with the wave vector $\mathbf{q}$ can be obtained by solving the following eigenvalue problem: \cite{dove}

\begin{equation}\label{eqn: hld}
\omega^2 (\mathbf{q}, \nu)\mathbf{e}(\mathbf{q}, \nu) = \mathbf{D}(\mathbf{q})\mathbf{e}(\mathbf{q}, \nu).
\end{equation}

\noindent Here, $\mathbf{D}(\mathbf{q})$ is the dynamical matrix, which is constructed using the equilibrium positions of the atoms in the unit cell and the harmonic force constants, $\Phi_{ij}^{\alpha\beta}$. The harmonic force constants are defined as

\begin{equation}\label{eqn: harmonic fc}
\Phi_{ij}^{\alpha\beta} = \frac{\partial^2 U}{\partial u_i^\alpha \partial u_j^\beta},
\end{equation}

\noindent where $U$ is the potential energy of the system, $\alpha$ and $\beta$ denote Cartesian directions (i.e., $\alpha,\beta = x, y, z$), $i$ and $j$ denote atoms in the supercell, and $u_i^\alpha$ is a small displacement of atom $i$ in direction $\alpha$. We calculate the harmonic force constants numerically using a central finite difference of DFT energies with respect to small perturbations of the equilibrium structure. The finite difference formulas are provided in Sec.~S2A of the Supplemental Material. \footnote{See Supplemental Material for information about ensemble lattice constants, ensemble thermal conductivities calculated the using ensemble lattice constants, silicon phonon dispersions (including degenerate modes), histograms of ensemble TA and LA sound speeds, mode-dependent Gr{\"u}neisen parameters, distributions fits to ensemble thermodynamic quantities, finite difference formulas for harmonic and third-order force constants, and a calculation of thermal conductivity where the force constants were calculated using finite differences of forces rather than of energies.} We calculate the force constants using the energies, as opposed to the atomic forces as is typically done, \cite{alan_jap} because the BEEF-vdW ensemble estimates uncertainty in the energy and not in the forces. We show in Sec.~S3 that obtaining the force constants from the energies or the forces yields the same phonon properties and thermal conductivity using the LDA XC functional.

The volumetric specific heat and group velocity in Eq.~\eqref{eqn: lattice k} can be calculated using the output of a harmonic lattice dynamics calculation. The total volumetric specific heat $c$ is given by \cite{alan_jap}

\begin{align}
c &= \frac{1}{V}\sum_{\mathbf{q},\nu} c(\mathbf{q},\nu) =\frac{1}{V}\sum_{\mathbf{q},\nu} \frac{\kb x^2 e^x}{(e^x-1)^2}, \label{eqn:specific heat}
\end{align}

\noindent where $V$ is the volume of the crystal and $x=\hbar \omega(\mathbf{q},\nu)/\kb T$, where $\hbar$ is the reduced Planck constant, $\kb$ is the Boltzmann constant, and $T$ is the temperature. To facilitate comparison with experimental values, we also calculate specific heat in J/kg-K using the conversion factor $V/(m\cdot n_q\cdot n_{basis})$, where $m$ is the atomic mass, $n_q$ is the number of phonon wave vectors, and $n_{basis}$ is the number of atoms in the unit cell. The group velocity is given by \cite{wang}

\begin{align}
\mathbf{v}_g(\mathbf{q},\nu) &= \frac{\partial \omega(\mathbf{q},\nu)}{\partial \mathbf{q}}=\frac{1}{2\omega(\mathbf{q}, \nu)}\left[\mathbf{e}^\dagger (\mathbf{q}, \nu)\frac{\partial \mathbf{D}(\mathbf{q})}{\partial \mathbf{q}}\mathbf{e}(\mathbf{q}, \nu) \right],\label{eqn:group velocity}
\end{align}

\noindent where the superscript $\dagger$ indicates a conjugate transpose. The group velocity is calculated by approximating the derivative of the dynamical matrix in Eq.~\eqref{eqn:group velocity} with a three-point central finite difference formula.

\subsubsection{\label{sec:ald}Anharmonic lattice dynamics and the Boltzmann transport equation}

Anharmonic lattice dynamics and BTE calculations are required to determine the intrinsic three-phonon scattering rates that are necessary to calculate the lifetimes in Eq.~\eqref{eqn: lattice k}. The intrinsic scattering rate for a three-phonon interaction is given by the Fermi golden rule, which requires as input harmonic phonon properties (Sec.~\ref{sec:hld}), the atomic masses, and the cubic force constants $\Psi_{ijk}^{\alpha\beta\gamma}$, which are defined as 

\begin{equation}\label{eqn: anharmonic fc}
\Psi_{ijk}^{\alpha\beta\gamma} = \frac{\partial^3 U}{\partial u_i^\alpha \partial u_j^\beta\partial u_k^\gamma}.
\end{equation}

\noindent The cubic force constants are calculated similarly to the harmonic force constants using a central finite difference formula on the energy that is presented in Sec.~S2B. Along with the harmonic quantities described in Sec.~\ref{sec:hld} and the cubic force constants, the phonon mode populations are needed to determine the lifetimes. The mode populations are calculated by solving the phonon BTE, which we do using an iterative approach. \cite{omini}

\subsection{\label{sec:computational detals}Computational details}

Self-consistent DFT calculations were performed with the real-space projector-augmented wave method \cite{blochl, kresse} as implemented in GPAW. \cite{gpaw1, gpaw2} We used the PBE, PBEsol, LDA, optPBE-vdW, and BEEF-vdW XC functionals. The BEEF-vdW XC functional was used with 2000 ensemble functionals for each calculation. Using more than 2000 functionals has been found to have little effect on the standard deviation of the ensemble energy values. \cite{beef, ahmad} We used a real-space grid spacing of $0.18$ $\A$. To calculate the harmonic force constants, we used a $3\times3\times3$ supercell consisting of 216 atoms with a $1\times1\times1$ electronic wave vector grid, while we used a $2\times2\times2$ supercell (64 atoms) and a $2\times2\times 2$ electronic wave vector grid for the cubic force constants. All energies were converged so that the variation between the final three iterations was at most $10^{-9}$ eV. 

To determine the zero-pressure lattice constants, energies were calculated for a series of strains. For each XC functional, five equally-spaced points between a maximum compression of $0.95$ times the experimental lattice constant of $5.430$ \AA \cite{icsd} and a maximum tension of $1.05$ times the experimental lattice constant were used to fit a third-order polynomial. A wider range of $0.85$ to $1.15$ times the experimental lattice constant with ten equally-spaced points was used for each member of the ensemble. The zero-pressure lattice constant corresponds to the minimum energy of the fitted polynomial. \cite{Alchagirov2003} 

Atomic displacements of $\pm 0.01$ $\A$ were applied to calculate the harmonic and cubic force constants using equations  in Sec.~S2. The harmonic and cubic force constant cutoffs correspond to the tenth and third nearest-neighbors (i.e., $1.5$ and $0.9$ lattice constants). A $24\times24\times24$ phonon wave vector grid was used to predict the harmonic phonon properties and thermal conductivity. This grid is based on the convergence testing of Jain and McGaughey. \cite{ankit} Translational invariance in the harmonic and cubic force constants was enforced using the Lagrangian approach presented by Li et al. \cite{li} 

\section{\label{sec:silicon}Results}

\subsection{\label{subsec:silicon overview}Overview}

We now apply the proposed framework to isotopically-pure silicon. A summary of the key results and relevant values from Jain and McGaughey, \cite{ankit} who used Quantum Espresso (QE) \cite{qe} for their DFT calculations, are provided in Table \ref{tbl: si results}. To ensure proper comparison to our values, we only include results where Jain and McGaughey used PAW pseudopotentials. The spreads reported for the BEEF-vdW calculations are the sample standard deviations of the ensemble predictions. 

\begingroup
\squeezetable
\begin{table}
\caption{\label{tbl: si results}Predicted lattice constant, transverse acoustic (TA) phonon frequency at the $\mathbf{X}$ point, [100] longitudinal acoustic (LA) sound speed, three-phonon phase space, average Gr{\"u}neisen parameter, and thermal conductivity (at $T=300$ K) of silicon using different XC functionals. The spreads reported for the BEEF-vdW calculations are the sample standard deviations of the BEEF-vdW ensemble predictions. Values in parentheses indicate the deviation of the quantity from the BEEF-vdW best-fit value, where $\sigma$ is the ensemble standard deviation.}% QE is all dfpt
\begin{ruledtabular}
\begin{tabular}{>{\RaggedLeft\arraybackslash}p{1.8cm}|>{\RaggedLeft\arraybackslash}p{1.15cm}|>{\RaggedLeft\arraybackslash}p{2.2cm}|>{\RaggedLeft\arraybackslash}p{2cm}|>{\RaggedLeft\arraybackslash}p{2.2cm}|>{\RaggedLeft\arraybackslash}p{2cm}|>{\RaggedLeft\arraybackslash}p{2cm}|>{\RaggedLeft\arraybackslash}p{2.0cm}}
XC Potential & DFT Package & Lattice constant, $a$ ($\A$) & TA frequency at $\mathbf{X}$-point (THz) & [100] LA sound speed (m/s) & Three-phonon phase space ($\times 10^{-3})$ & Average Gr{\"u}neisen parameter, $\bar{\gamma}$ & Thermal conductivity at $T = 300$ K (W/m-K)  \\
\hline
 BEEF-vdW & GPAW & $5.479\pm0.077$ & $4.82\pm 0.57$ & $8564\pm 162$ & $1.15 \pm 0.05$ & $0.92\pm 0.14$ &  $171\pm 24$\\
  optPBE-vdW & GPAW & 5.504 ($+0.32\sigma$) & 4.77 ($-0.09\sigma$) & 8475 ($-0.55\sigma$)&  1.16 ($+0.20\sigma$)& 0.92  ($0.00\sigma$)&  165 ($-0.25\sigma$)\\
LDA & GPAW & 5.408 ($-0.92\sigma$) & 3.97 ($-1.49\sigma$) & 8388 ($-1.09\sigma$) & 1.23 ($+1.60\sigma$)& 1.16 ($+1.71\sigma$)&  122 ($-2.04\sigma$) \\
  & QE\cite{ankit} & 5.400 ($-1.03\sigma$)&  & 8340 ($-1.38\sigma$)&  & 1.11 ($+1.36\sigma$)& 142 ($-1.21\sigma$) \\
PBE & GPAW & 5.478 ($-0.01\sigma$) & 4.58 ($-0.42\sigma$)& 8512 ($-0.32\sigma$)& 1.19 ($+0.80\sigma$)& 0.96 ($+0.29\sigma$)& 154 ($-0.71\sigma$) \\
  & QE\cite{ankit} & 5.466 ($-0.17\sigma$) &  & 7830 ($-4.53\sigma$)&  & 1.03 ($+0.79\sigma$)& 145 ($-1.08\sigma$) \\
PBEsol & GPAW & 5.442 ($-0.48\sigma$) & 4.03 ($-1.39\sigma$) & 8406 ($-0.97\sigma$)& 1.23  ($+1.60\sigma$)& 1.10 ($+1.29\sigma$)& 128 ($-1.79\sigma$) \\
 & QE\cite{ankit} & 5.430 ($-0.64\sigma$) & 4.04 ($-1.37\sigma$)& 8320 ($-1.51\sigma$)&   &  1.11 ($+1.36\sigma$)& 137 ($-1.42\sigma$) \\
Experiment &  & 5.430\footnotemark[1] ($-0.64\sigma$) & 4.48\footnotemark[2] ($-0.60\sigma$) & 8430\footnotemark[3]  ($-0.83\sigma$) &  & &  153\footnotemark[2] ($-0.75\sigma$)\\
\end{tabular}
\end{ruledtabular}
\footnotetext[1]{Ref.~\citenum{icsd}}
\footnotetext[2]{Ref.~\citenum{si_expt}}
\footnotetext[3]{Ref. \citenum{si_expt_sound_speed}}
\end{table}
\endgroup

%Experiment &  & 5.430 \footnote{Reference \cite{icsd}} ($-0.64\sigma$) & 4.48 \footnote{Reference \cite{si_expt}} ($-0.60\sigma$) & 8430 \footnote{Reference \cite{si_expt_sound_speed}} ($-0.83\sigma$) &  & &  153 \footnote{Reference \cite{si_expt}} ($-0.75\sigma$)\\

\subsection{\label{subsec:silicon lat}Lattice constant}

The lattice constant is accurately predicted by all XC functionals tested, with a maximum deviation of 1.36\% from the experimental value ($5.430$ $\A$) \cite{icsd} from optPBE-vdW ($5.504$ $\A$). LDA is the only XC functional that under-predicts the lattice constant, while all other functionals (PBE, PBEsol, optPBE-vdW, and BEEF-vdW) over-predict it. The same trend was observed by Jain and McGaughey \cite{ankit} and is consistent with previous observations that LDA tends to overestimate binding strength. \cite{haas} The ensemble lattice constants are determined by fitting and minimizing an equation of state with respect to energy for each BEEF-vdW ensemble functional. A histogram of the results is provided in Fig.~S1(a). All predicted self-consistent lattice constants and the experimental value are bounded to within one ensemble standard deviation ($\sigma=0.077$ $\A$) of the BEEF-vdW best-fit value of $5.479$ $\A$, with the exception of the LDA lattice constant from Jain and McGaughey ($-1.03\sigma$). 

There is an ambiguity in the choice of the lattice constant to be used in the ensemble lattice dynamics calculations. There are two possible approaches: (i) Use the lattice constant from the BEEF-vdW best-fit XC functional for all calculations because the ensemble force constants are calculated at this lattice constant. (ii) For each member of the ensemble, use the lattice constant determined from that member's equation of state. While we believe that both choices are reasonable, we chose to use (i), the BEEF-vdW best-fit lattice constant, because it is consistent with the ensemble force constant calculations. The effect of this choice on the ensemble thermal conductivity predictions is explored in Sec.~\ref{subsec:silicon k}.

\subsection{\label{subsec:silicon harmonic}Phonon dispersion, sound speed, and specific heat}

\begin{figure}
\begin{tabular}{c}
  \includegraphics[width=3.4 in]{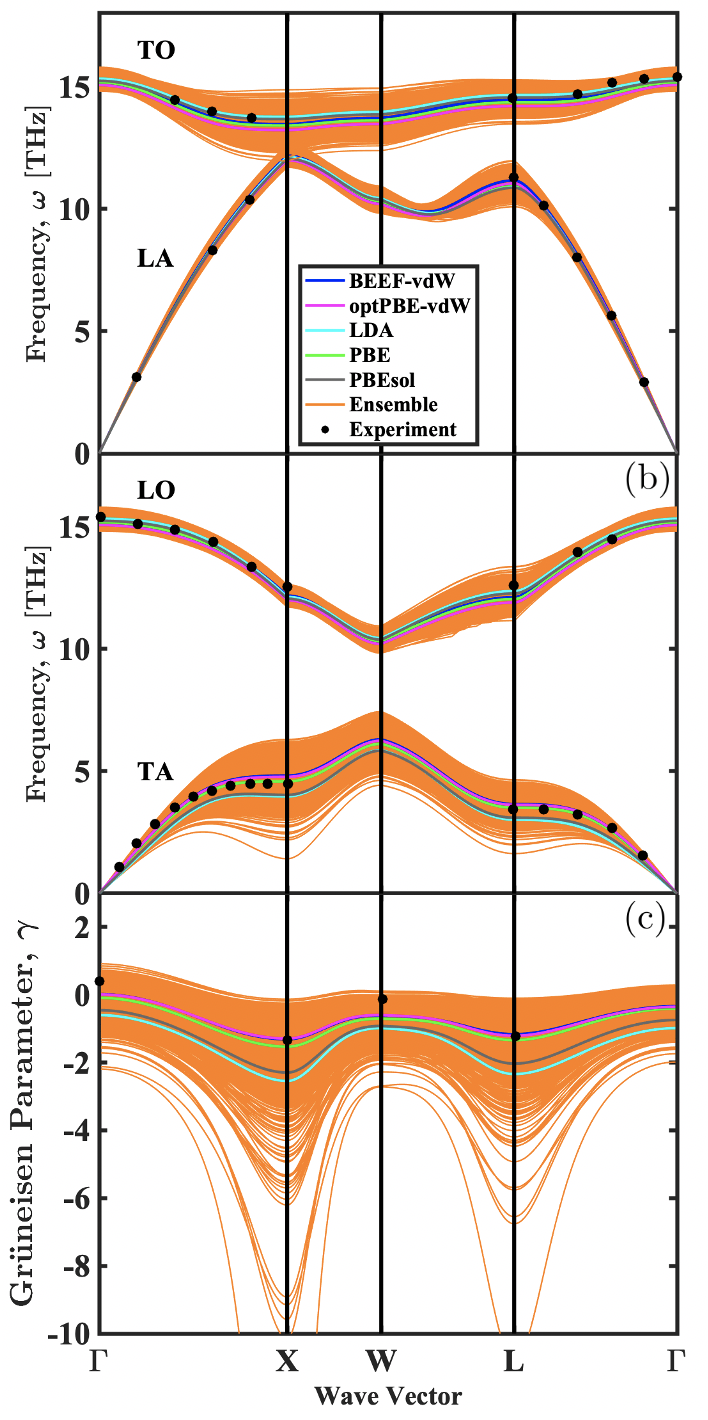}
\end{tabular}
\caption{Silicon phonon dispersion relations for the (a) LA and TO branches and (b) TA and LO branches along high-symmetry directions. (c) Gr{\"u}neisen parameters for the TA branch plotted along the same high-symmetry directions. The cut-off Gr{\"u}neisen parameter curve reaches $-25$ at the $\mathbf{X}$-point. 
%DFT calculations were performed with the BEEF-vdW (blue), LDA (cyan), PBE (green), optPBE-vdW (magenta), PBEsol (grey), and ensemble (orange) functionals. 
Experimental dispersion values are from Nilsson and Nelin\cite{nilsson} and Gr{\"u}neisen parameter values are from Madelung et al. \cite{si_grun} Each curve traces along the $\mathbf{\Gamma}$ $(0, 0, 0)$, $\mathbf{X}$ $(1,0,0)$, $\mathbf{W}$ $(1, 0.5, 0)$, and $\mathbf{L}$ $(0.5, 0.5, 0.5)$ reduced wave vector points in the Brillouin zone.}
\label{fig: dispersions}
\end{figure}

Predicted and experimental \cite{nilsson} phonon dispersion relations on the $\mathbf{\Gamma}-\mathbf{X}-\mathbf{W}-\mathbf{L}-\mathbf{\Gamma}$ loop are plotted in Figs.~\ref{fig: dispersions}(a) and \ref{fig: dispersions}(b). The transverse branches are degenerate on $\mathbf{\Gamma}-\mathbf{X}-\mathbf{W}$ and $\mathbf{\Gamma}-\mathbf{L}$. On $\mathbf{W}-\mathbf{L}$, for clarity, only the lower frequency transverse branch is plotted. All branches, including the two excluded ones, are plotted in Figs.~S2(a)-S2(f). The ensemble bounds the experimental and self-consistent DFT dispersions. The greatest spread amongst the self-consistent DFT dispersions is found in the transverse acoustic (TA) and longitudinal optical (LO) branches at the Brillouin zone edge $\mathbf{X}$-point. This behavior is mirrored in the ensemble dispersions. As noted in Table~\ref{tbl: si results}, the TA branch frequency has a standard deviation of $0.57$ THz at the $\mathbf{X}$-point, compared to a standard deviation of only $0.14$ THz for the longitudinal acoustic (LA) branch at that point. Some of this TA branch spread is due to some ensemble members decreasing in frequency near the $\mathbf{X}$-point, a result that contradicts experimental observations. \cite{nilsson} Previous authors have also noted difficulty in using lattice dynamics to model the TA branch in silicon and germanium, \cite{richter, herman} which has been ascribed to the sensitivity to the number of neighbor shells included in the calculation. \cite{alan_jap, esfarjani, mazur} The ensemble results demonstrate that the TA branch is also sensitive to the force constants. 

The sound speed is calculated using Eq.~\eqref{eqn:group velocity} near the $\mathbf{\Gamma}$-point for the LA branch in the [100] (i.e., $\mathbf{\Gamma}-\mathbf{X}$) direction. The experimental value of $8430$ m/s and all self-consistent DFT predictions are bounded to within two ensemble standard deviations of the BEEF-vdW best-fit value of $8564$ m/s with the exception of the PBE value from Jain and McGaughey \cite{ankit} ($-4.53\sigma$). Histograms of the sound speed of the TA and LA branches are shown in Figs.~S3 and S4.

\begin{figure*}
\includegraphics[width=3.4in]{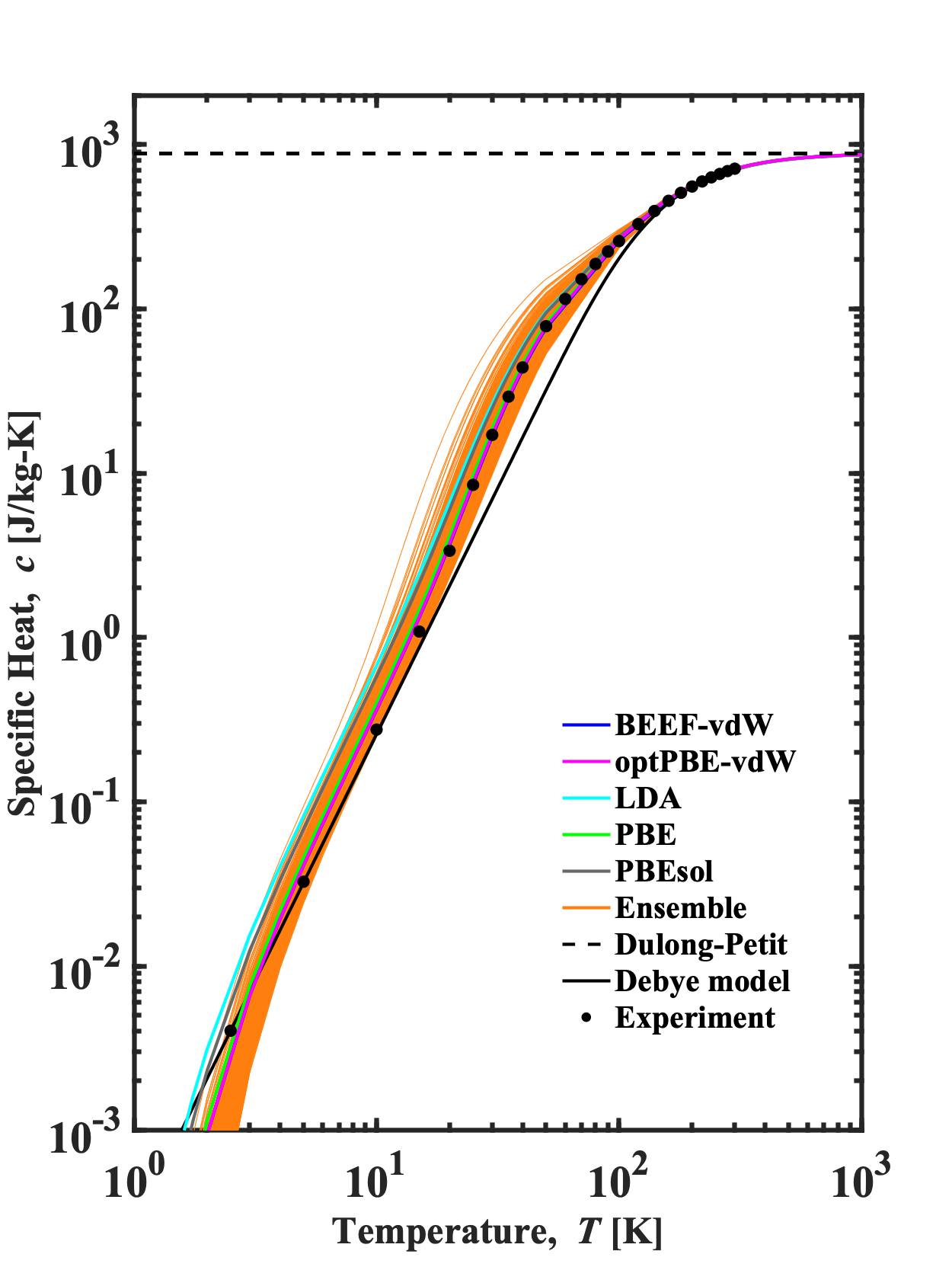}
\caption{DFT and Debye model predictions of silicon specific heat. The Dulong-Petit limit is shown as a dotted line at $888$ J/kg-K. The experimental values are from Flubacher et al. \cite{flubacher}}
\label{fig: cp}
\end{figure*}

Specific heat values in units of J/kg-K are plotted in Fig.~\ref{fig: cp} as a function of temperature between $1$ and $1000$ K. Experimental values from Flubacher et al. \cite{flubacher} are shown for comparison. The experimental data are bounded by the ensemble over the entire temperature range. The largest spread in the ensemble is $12$ J/kg-K at a temperature of $50$ K, which is 15\% of the BEEF-vdW best-fit value of 76 J/kg-K at that temperature. As temperature is increased, $x=\hbar\omega/\kb T$ gets smaller, such that the differences in frequencies predicted by each ensemble are suppressed. This effect is reflected in the narrowing of the ensemble predictions above a temperature of 100 K. As the temperature approaches 1000 K, all ensemble and DFT self-consistent predictions approach the Dulong-Petit limit of $3\kb/m = 888$ J/kg-K. \footnote{Guan et al. \cite{guan} used the BEEF-vdW ensemble to calculate the constant pressure specific heat $c_p$ of eight materials of various crystal structures from \cite{Guan2017}

\[ c_p = c +T\alpha^2 B V. \]

\noindent Here, $c$ is given by Eq.~\eqref{eqn:specific heat}, $\alpha$ is the thermal expansion coefficient, and $B$ is the bulk modulus. They find an increase in the uncertainty of their $c_p$ predictions with increasing temperature because of a corresponding increase in uncertainty in their $\alpha$ prediction.}

We also include the specific heat predicted using the Debye model,

\begin{equation}
    c_\text{Debye} = \frac{9\kb}{m} \left(\frac{T}{\theta}\right)^3 \int_0^{\theta/T} \frac{x^4 e^x}{(e^x-1)^2}dx,
\end{equation}

\noindent where $\theta=645$ K is the Debye temperature for silicon. \cite{si_expt} The Debye model prediction is worse than any of the ensemble predictions in the $30$ to $100$ K range, but is as accurate as any self-consistent calculation at temperatures below $10$ K and above $100$ K. The agreement at low temperatures is because the assumption of a linear dispersion relation for all phonons in the Debye model is most accurate at temperatures much lower than the Debye temperature. \cite{si_expt} 

\subsection{\label{subsec:silicon gruneisen}Gr{\"u}neisen parameter and thermal expansion coefficient}

The mode-dependent Gr{\"u}neisen parameters quantify the effect of crystal volume change on the phonon frequencies and are a measure of anharmonicity. \cite{ankit} We calculated them using the cubic force constants through Eq.~(2) from Fabian and Allen. \cite{fabian} The results for the TA branch are plotted in Fig.~\ref{fig: dispersions}(c) on the $\mathbf{\Gamma}-\mathbf{X}-\mathbf{W}-\mathbf{L}-\mathbf{\Gamma}$ loop. The remaining branches are plotted in Figs.~S5(a)-S5(f). The TA branch at the $\mathbf{X}$-point has the largest spread of any of the modes, with a standard deviation of $1.00$. At the $\mathbf{X}$-point, the largest deviation of any self-consistent prediction from the BEEF-vdW value of $-1.31$ is the LDA value of $-2.54$.

The average Gr{\"u}neisen parameter $\bar{\gamma}$ can be calculated as a specific heat-weighted average of mode Gr{\"u}neisen parameters from

\begin{equation}
\bar{\gamma} = \frac{\sum_{\mathbf{q}, \nu} c(\mathbf{q},\nu)|\gamma(\mathbf{q},\nu)|}{\sum_{\mathbf{q}, \nu} c(\mathbf{q},\nu)}.
\end{equation}

\noindent The average Gr{\"u}neisen parameter predictions are reported in Table \ref{tbl: si results}. BEEF-vdW and optPBE-vdW yield the lowest average value (0.92) and LDA yields the highest (1.16). All self-consistent predictions are bounded to within two standard deviations of the BEEF-vdW value. The differences in predictions of anharmonicity in both the self-consistent and ensemble calculations are correlated with the predicted thermal conductivity, an effect that we explore in Sec.~\ref{subsec:silicon discussion}.

The mode-dependent Gr{\"u}neisen parameters can be used to calculate the thermal expansion coefficient (TEC). \cite{ritz} The ensemble TEC values for silicon are compared to the values from Guan et al.\cite{guan} in Sec.~S1E. While Guan et al. did not perform calculations for silicon, the coefficient of variation (COV $=$ standard deviation/mean) for the silicon TEC ensemble is $0.40$, consistent with the range of values they report (0.26 to 0.75). In contrast to our negatively-skewed silicon TEC distribution, however, all TEC distributions reported by Guan et al. have a positive skew.

\subsection{\label{subsec:silicon k}Thermal conductivity}

\begin{figure*}
  \includegraphics[width=3.4in]{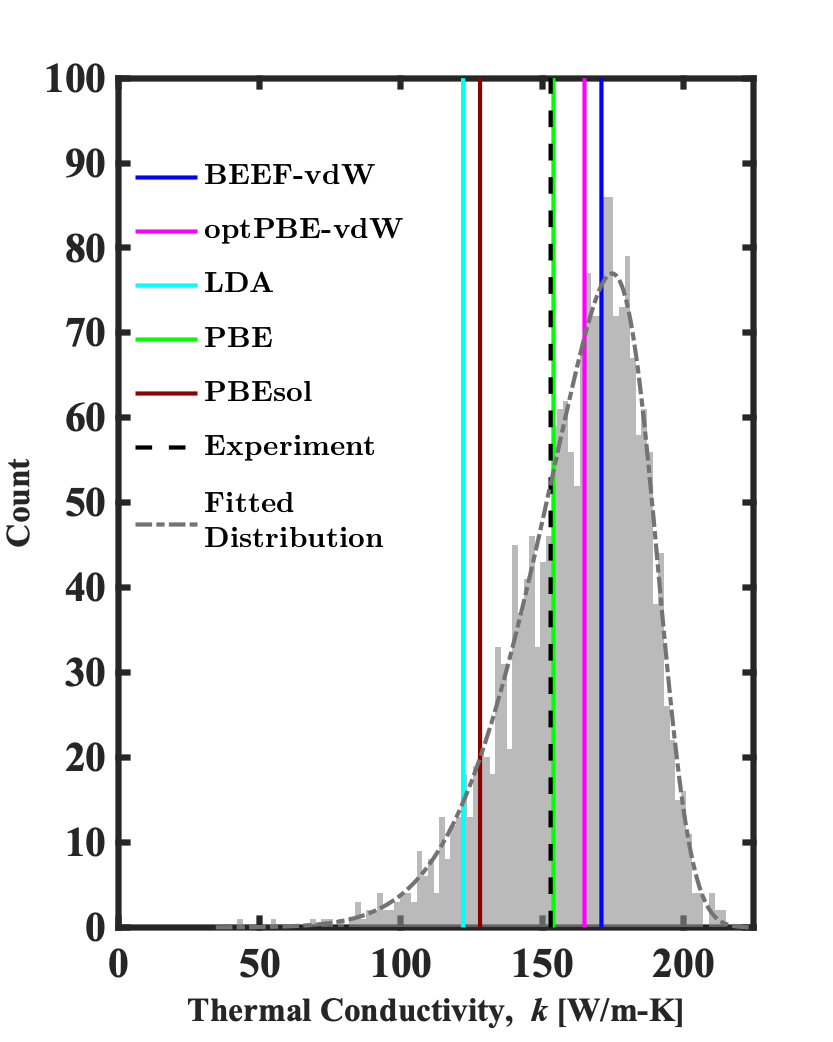}
% \multicolumn{2}{c}{\includegraphics[width=65mm]{it} }\\
% \multicolumn{2}{c}{(e) fifth}
\caption{DFT predictions of silicon thermal conductivity at $T=300$ K, with the BEEF-vdW ensemble shown in gray. The lightest (middle) shade of gray indicates a range of $\pm\sigma$ ($\pm2\sigma$) around the BEEF-vdW best-fit value. The experimental value is from Inyushkin et al. \cite{si_expt} The overlaid distribution is a skewed normal distribution with mean $190$ W/m-K, standard deviation $36$ W/m-K, and skewness $-3.7$.}
\label{fig: ens_k}
\end{figure*}

The thermal conductivity results are plotted in Fig.~\ref{fig: ens_k}. The BEEF-vdW best-fit value is 171 W/m-K and the ensemble has a standard deviation of 24 W/m-K. The ensemble distribution is not symmetrical, with a longer tail to the left of its mean. A majority of ensemble functionals (1165 out of 2000) predict a lower thermal conductivity than the BEEF-vdW best-fit value. Following the procedure outlined by Guan et al. \cite{guan} for fitting distributions based on the Cramer von Mises goodness of fit test, \cite{anderson1952} the distribution is best described ($p$-value $=0.94$) by a skewed normal distribution, which is also plotted in Fig.~\ref{fig: ens_k}, with a mean of $190$ W/m-K, a standard deviation of $36$ W/m-K, and a skewness of $-3.7$. Additional distribution fits for other ensemble quantities are presented in Sec.~S1E. The BEEF-vdW best-fit prediction is higher than the experimental value of 153 W/m-K. \cite{si_expt} An overestimation is reasonable because the prediction framework does not account for isotope, phonon-boundary, phonon-defect, or four-phonon scattering, all of which reduce thermal conductivity. 

The BEEF-vdW ensemble bounds nearly all self-consistent DFT predictions, including those from Jain and McGaughey, \cite{ankit} to within two ensemble standard deviations of the BEEF-vdW best-fit value. The largest discrepancy is the $122$ W/m-K thermal conductivity prediction from GPAW LDA ($-2.04\sigma$ from the BEEF-vdW value). There is no self-consistent thermal conductivity value in Table \ref{tbl: si results} greater than the BEEF-vdW best-fit functional prediction. It is noteworthy that the optPBE-vdW value of $165$ W/m-K is closest to the BEEF-vdW value. Parks et al. \cite{beef_molecules} found that XC functionals that include vdW correlations, such as BEEF-vdW and optPBE-vdW, tend to predict higher vibrational frequencies for molecules and molecular complexes compared to GGA-level counterparts that do not include vdW correlations. A similar result is observed here in the BEEF-vdW and optPBE-vdW phonon dispersions in Figs.~\ref{fig: dispersions}(a) and \ref{fig: dispersions}(b). BEEF-vdW and optPBE-vdW predict the highest $\mathbf{X}$-point frequencies for both the TA and LA branches, which results in higher group velocities and thus higher thermal conductivity.

% \footnote{This under-prediction is consistent with the phonon properties predicted by the LDA functional, such as low TA and LA group velocities, but is not consistent with reported LDA thermal conductivities () \cite{ankit, alan_jap,lindsay, esfarjani, alloy, shengbte} but is consistent with other properties predicted by this functional, such as low TA and LA group velocities. Another possible interpretation is that the calculation is not be optimized with respect to computational parameters in either the lattice dynamics or DFT workflows. This is unlikely, as the lattice dynamics parameters were based on Jain and McGaughey's careful convergence testing, \cite{ankit} and the $0.18$ $\A$ grid spacing we used is equivalent to a $85$ Ry plane-wave cutoff in a plane-wave code, \cite{briggs} which should be sufficiently high cutoff to describe silicon. \cite{ankit} It should also be noted that the same parameters give reasonable results for other XC functionals.} 
 
\begin{figure*}
  \includegraphics[width=3.4in]{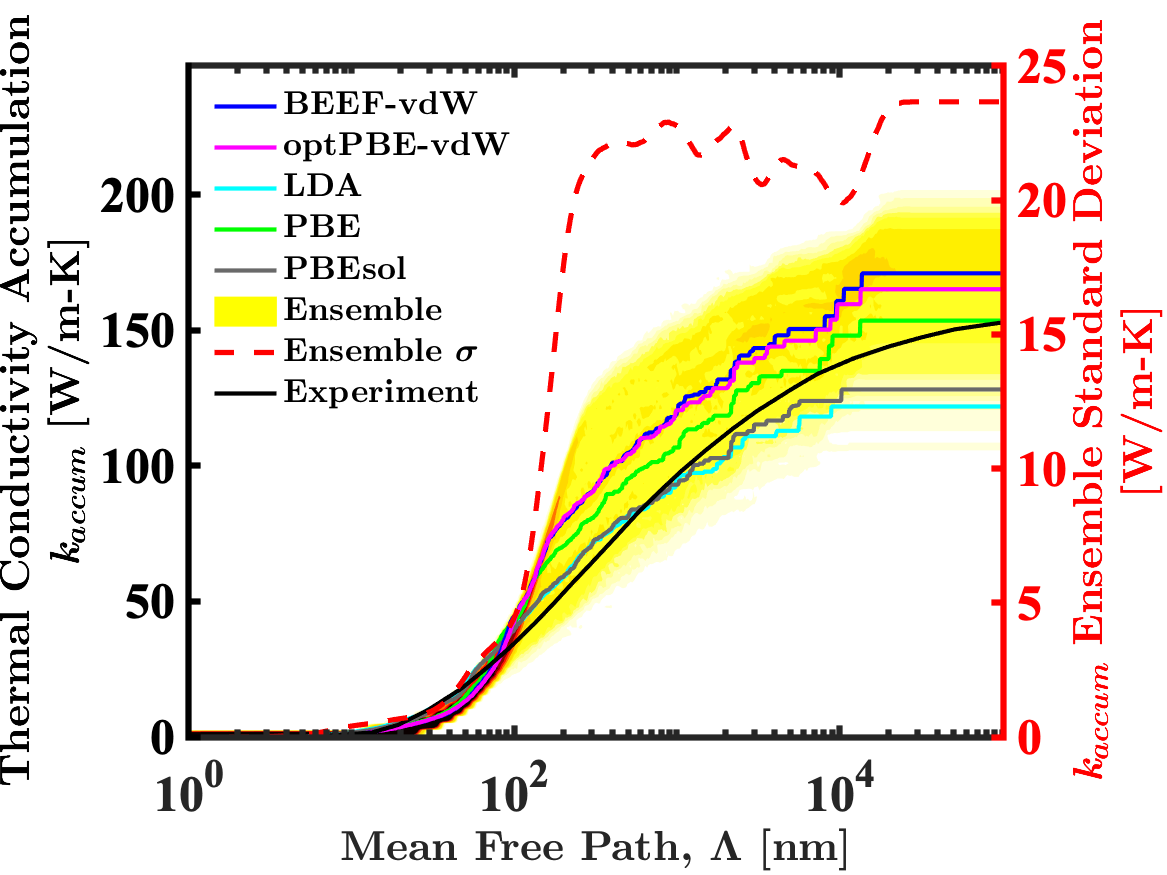}
% \multicolumn{2}{c}{\includegraphics[width=65mm]{it} }\\
% \multicolumn{2}{c}{(e) fifth}
\caption{Self-consistent, ensemble, and experimental \cite{cuffe} thermal conductivity accumulation functions of silicon at a temperature of $300$ K (left vertical axis). The red dotted line indicates the standard deviation of the ensemble thermal conductivity accumulation (right vertical axis).}
\label{fig: ens_accum}
\end{figure*}

The thermal conductivity accumulation function $k_{accum}(\Lambda)$ provides the contribution to thermal conductivity of phonons having mean free paths (MFP) less than $\Lambda$, where $\Lambda(\mathbf{q}, \nu)=|\mathbf{v}_{g}(\mathbf{q}, \nu)|\tau_l (\mathbf{q}, \nu)$ for heat flow in the $l$-direction. \cite{alan_jap} The ensemble results are plotted in Fig.~\ref{fig: ens_accum} as a heat map, with darker colors indicating a greater fraction of the ensemble predictions. The standard deviation of the ensemble thermal conductivity accumulation is also plotted. The experimental curve was determined by Cuffe et al.,\cite{cuffe} who used a transient thermal grating technique to measure the thermal conductivity of single-crystal silicon membranes of varying thickness.

All functionals indicate that phonons with MFPs shorter than $10$ nm do not contribute to thermal conductivity. The ensemble accumulation functions spread widely between MFPs of $100$ to $300$ nm, a trend that is reflected in the sharp increase of the ensemble standard deviation in this range. The spread then remains uniform up to $10^{4}$ nm, the longest MFP considered. This result suggests that ensemble members that predict a high thermal conductivity overpredict the contributions of phonons with MFPs between $100$ and $300$ nm. This interpretation is consistent with the GPAW BEEF-vdW (171 W/m-K) and optPBE-vdW (165 W/m-K) predictions and with the findings of Jain and McGaughey, \cite{ankit} who predicted a silicon thermal conductivity of $172$ W/m-K with the BLYP XC functional and attributed it to an overprediction of the contributions of $\sim 100$ nm MFP phonons. The experimental accumulation is bounded by the ensemble and has a similar slope compared to the self-consistent predictions, indicating agreement in the relative contributions of phonons with the displayed range of MFPs.

In Sec.~\ref{subsec:silicon lat}, we noted an ambiguity in the choice of the lattice constant for the ensemble lattice dynamics and BTE calculations. The thermal conductivities plotted in Fig.~\ref{fig: ens_k} are calculated using the BEEF-vdW lattice constant. Choosing instead to use the ensemble lattice constants changes any individual thermal conductivity by at most $7$ W/m-K and has no effect on the ensemble thermal conductivity standard deviation. A histogram of thermal conductivity values calculated using the ensemble lattice constants is shown in Fig.~S1(b). Based on the von Mises goodness of fit test ($p$-value $=0.97$), this distribution is also best described by a skewed normal distribution, with mean of $192$ W/m-K, a standard deviation of $38$ W/m-K, and a skewness of $-4.4$. This ensemble and its fitted distribution are nearly identical to the results obtained using the BEEF-vdW lattice constant shown in Fig.~\ref{fig: ens_k}.

\subsection{\label{subsec:silicon discussion}Thermal conductivity correlation analysis}

\begin{figure}[h]
\begin{tabular}{ccc}
(a) & (b) & (c) \\[-3pt]
  \includegraphics[width=2.0 in]{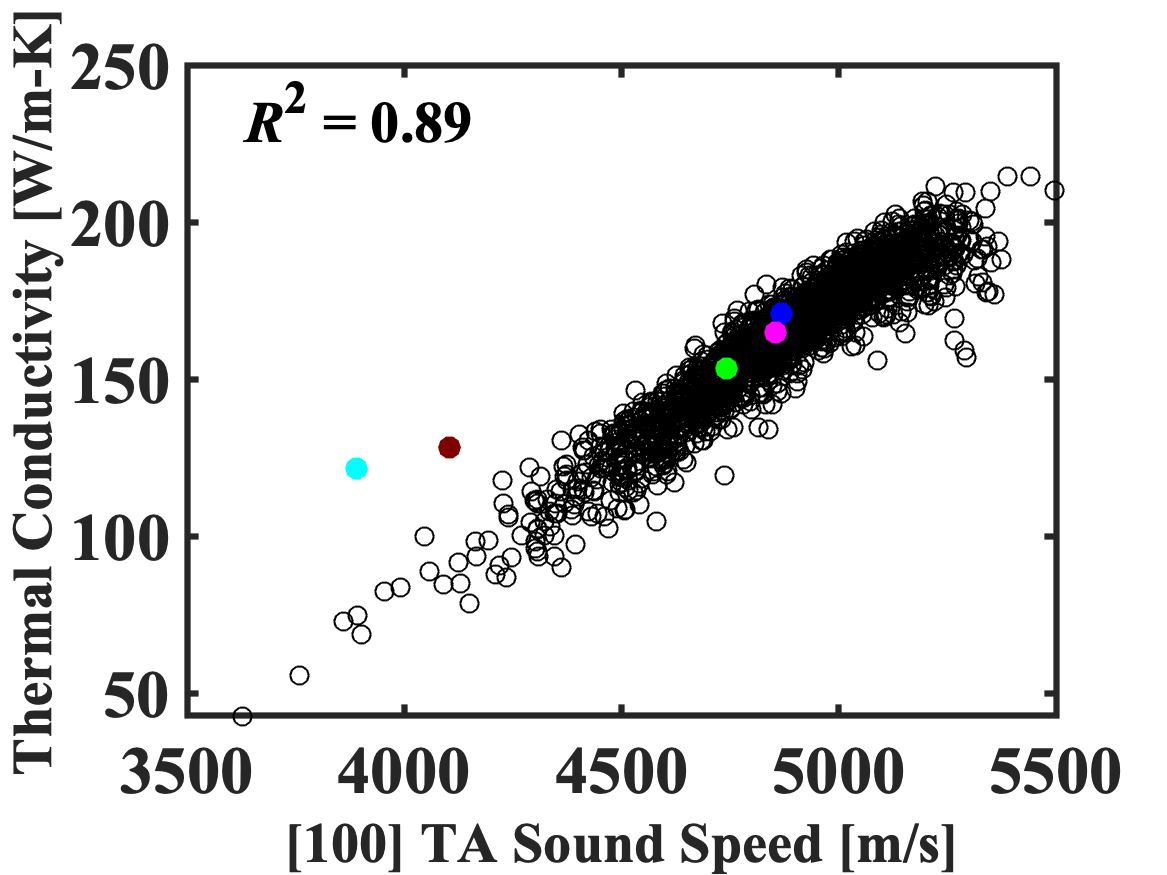} & \includegraphics[width=2.0 in]{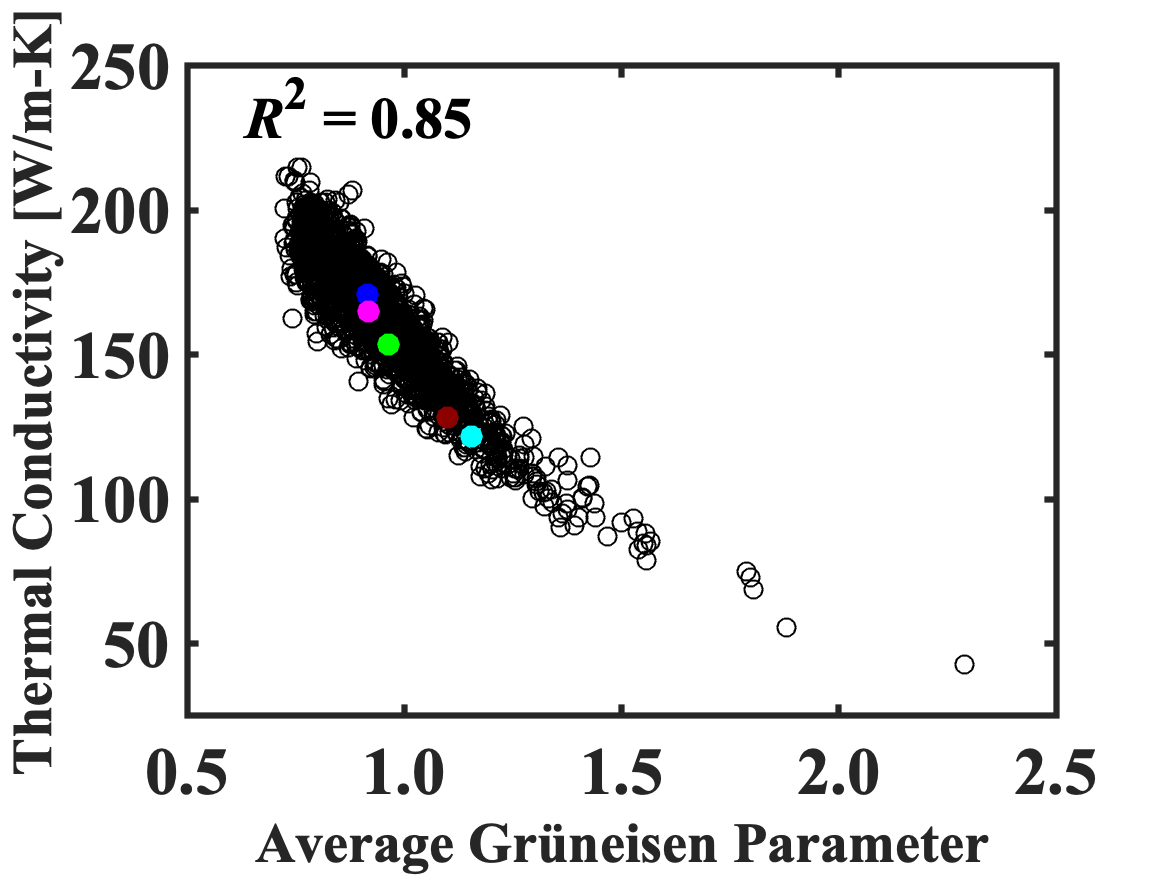}  &  \includegraphics[width=2.0 in]{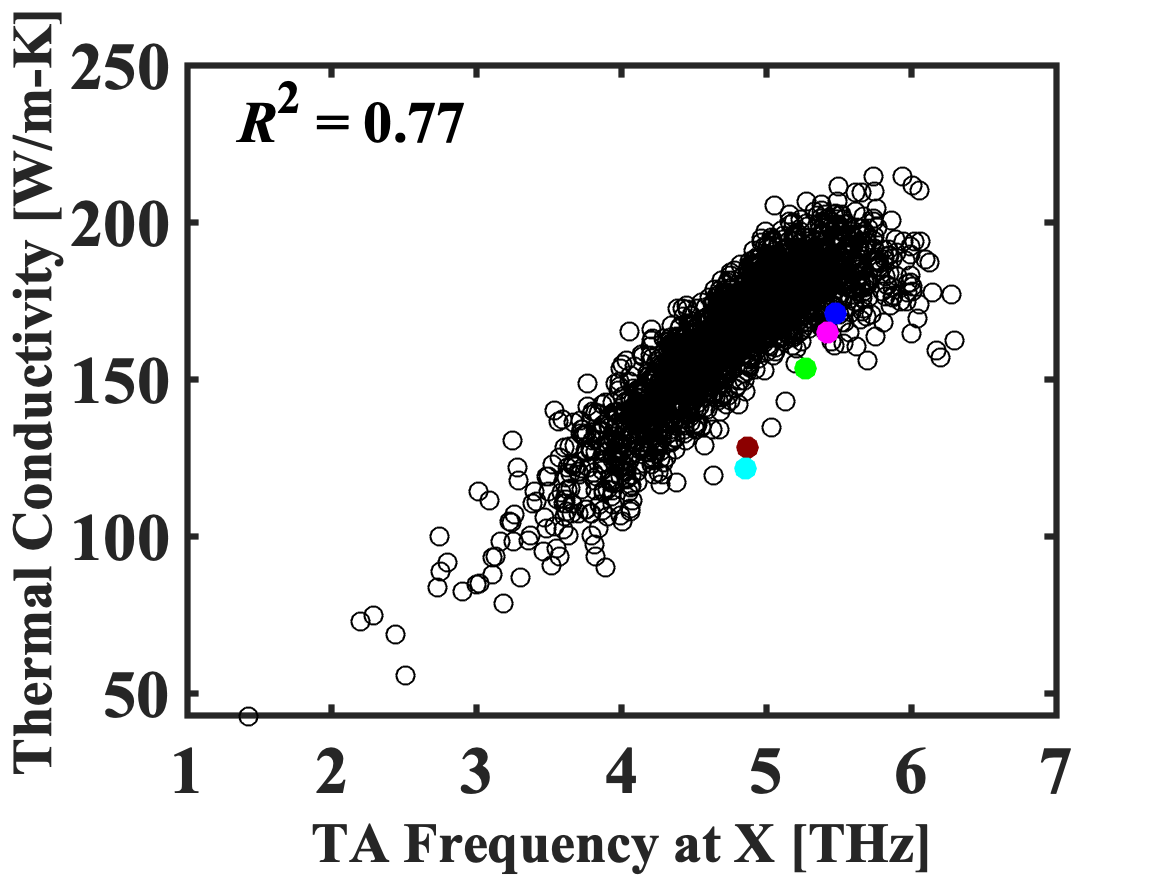} \\
     (d) & (e) & (f) \\[-3pt] 
    \includegraphics[width=2.0 in]{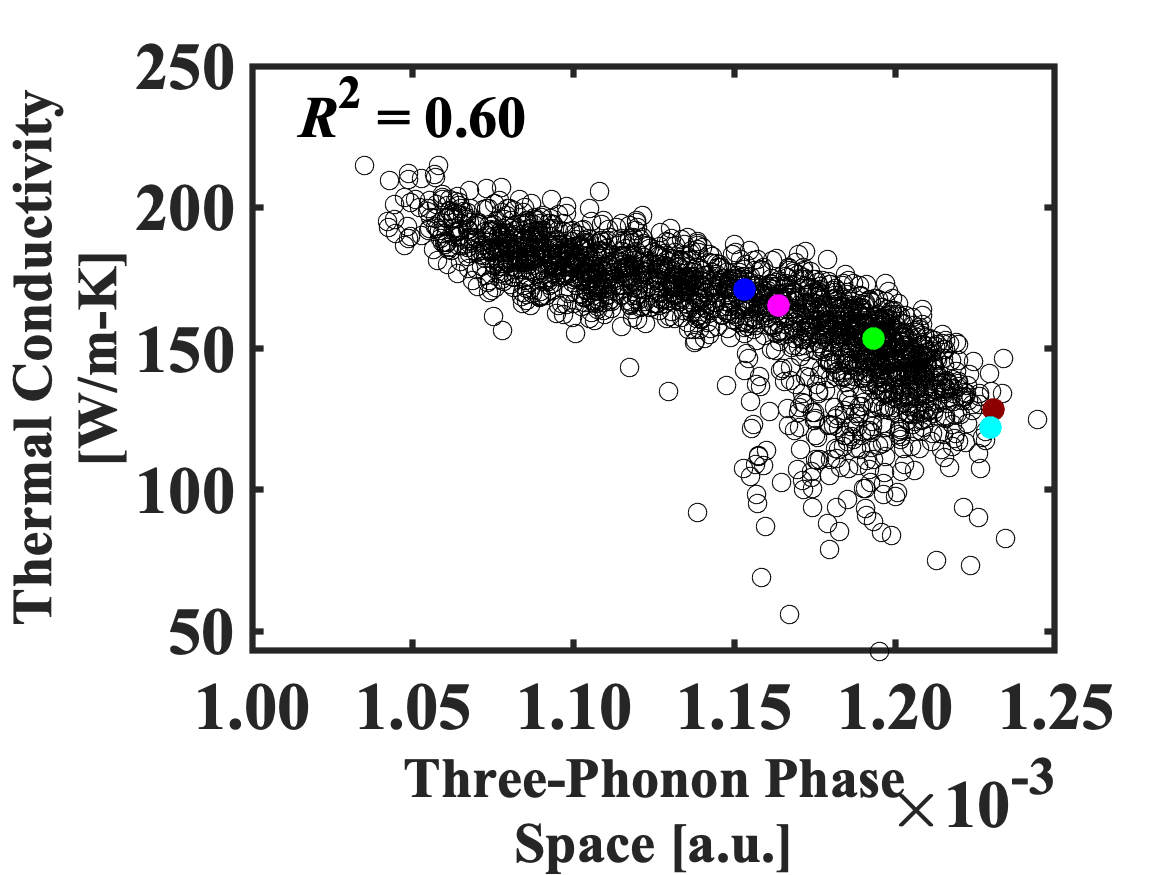} & \includegraphics[width=2.0 in]{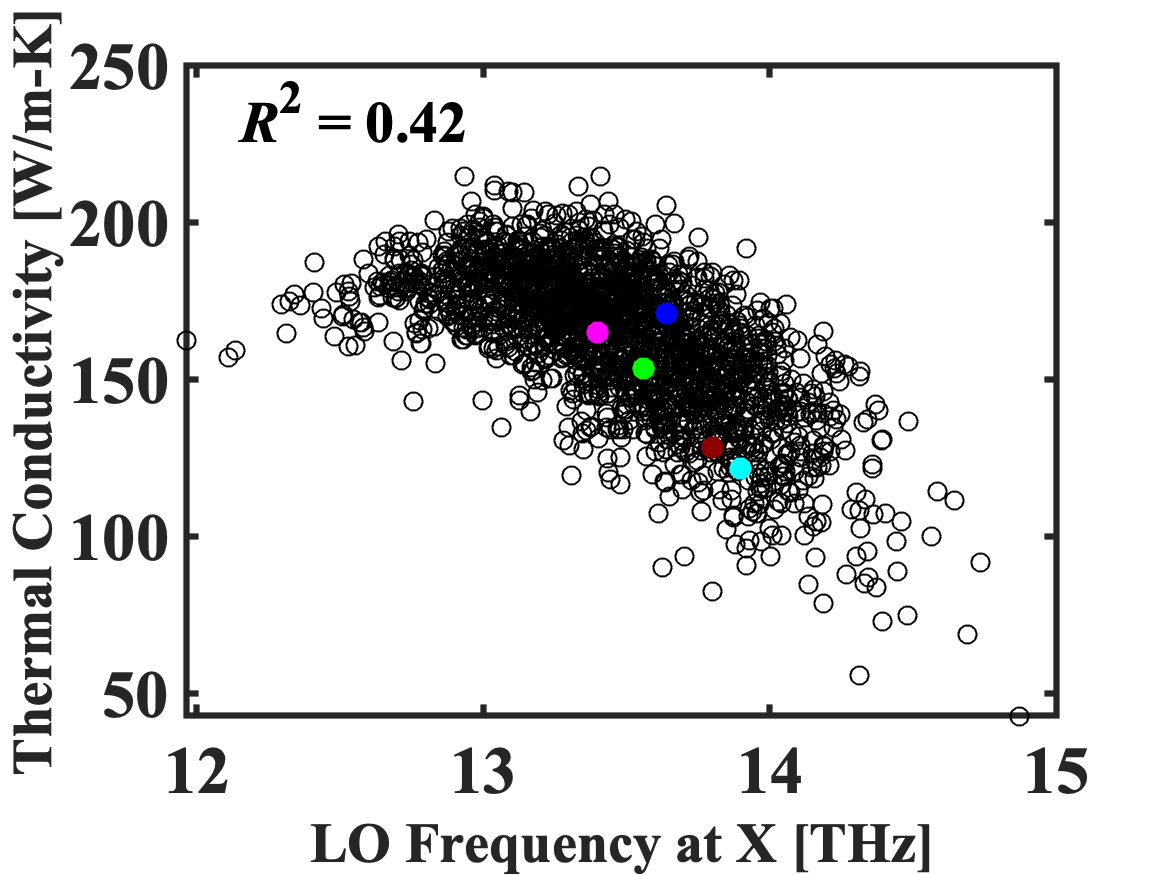} & \includegraphics[width=2.0 in]{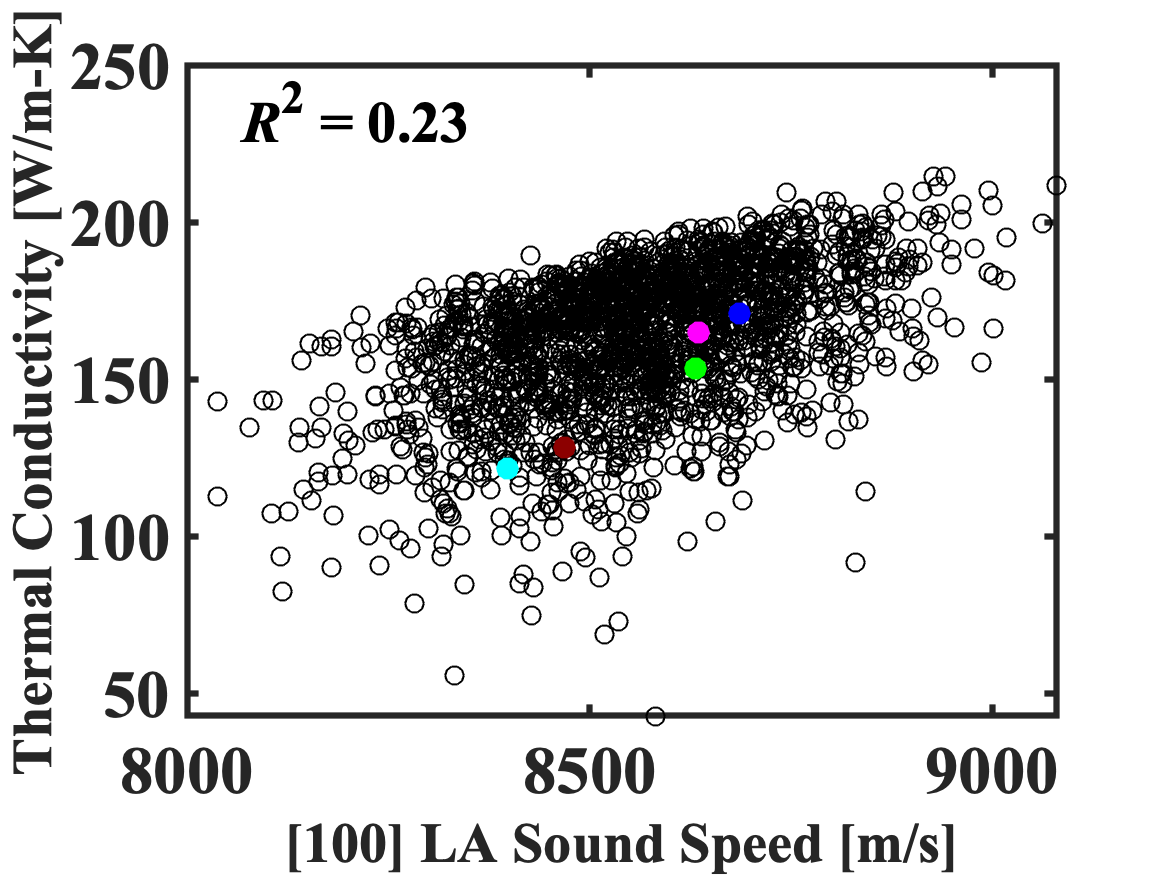} \\
(g) & (h) &  \\[-3pt]
   \includegraphics[width=2.0 in]{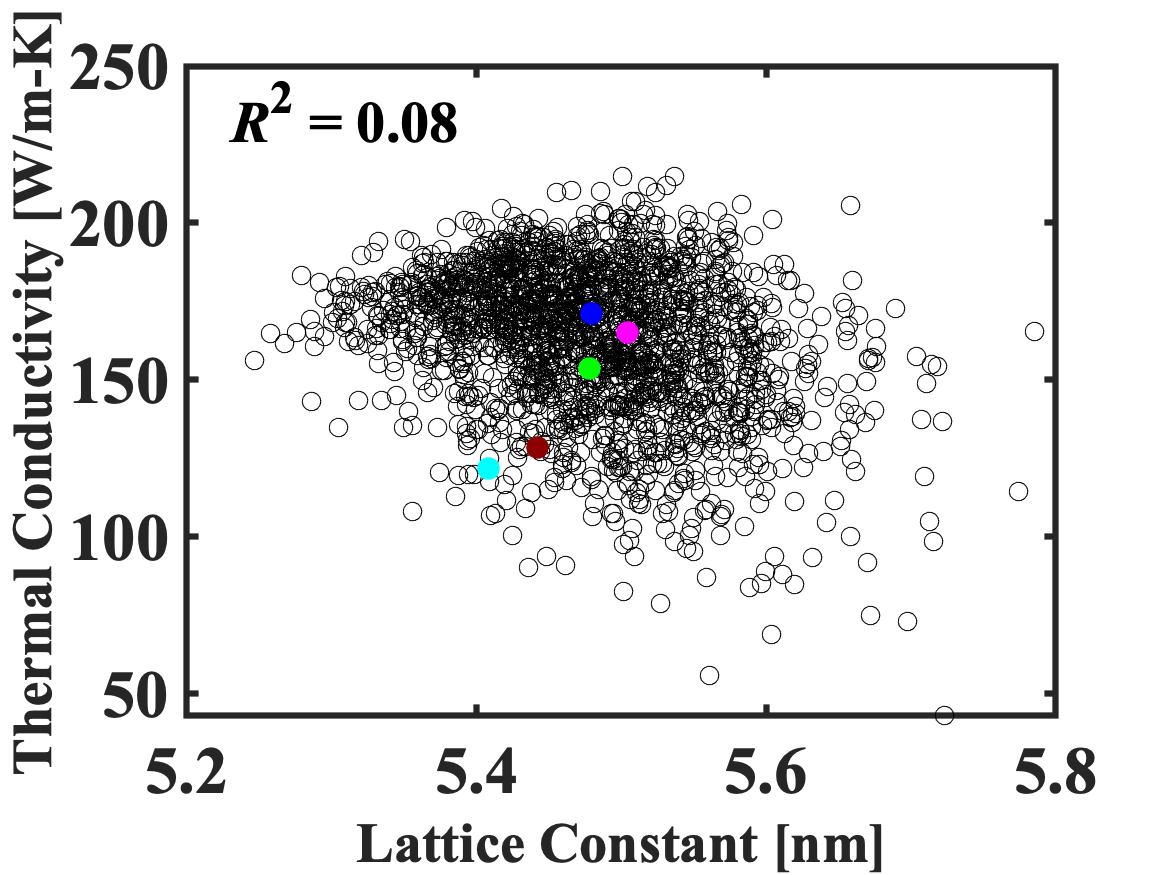} & \includegraphics[width=2.0 in]{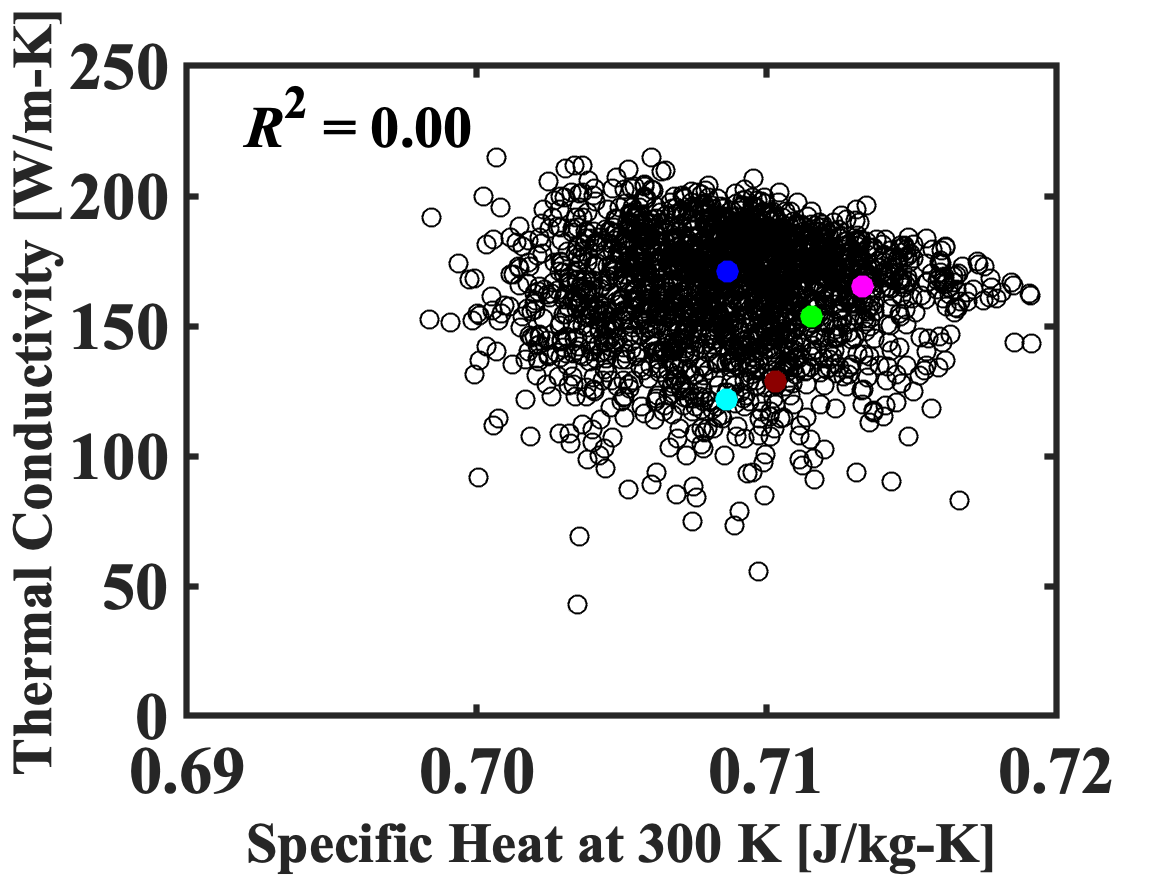} &  \includegraphics[width=1.0 in]{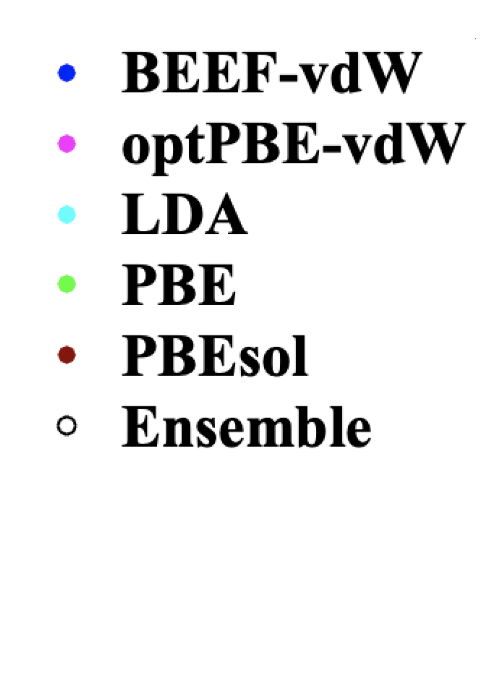} \\
\end{tabular}
\caption{Correlation of the BEEF-vdW ensemble predictions of thermal conductivity at $T=300$ K with other phonon and structural properties. The panels are presented in decreasing order of $R^2$. The quantities are (a) [100] TA sound speed, (b) average Gr{\"u}neisen parameter, (c) $\mathbf{X}$-point TA frequency, (d) three-phonon phase space, (e) $\mathbf{X}$-point LO frequency, (f) [100] LA sound speed, (g) lattice constant, and (h) specific heat at $T=300$ K. Self-consistent predictions are denoted with colored markers.}
\label{fig: scatter plots}
\end{figure}

We now examine how the spread in the ensemble thermal conductivities is related to the spreads in other ensemble quantities. The results are shown in Figs.~\ref{fig: scatter plots}(a)-\ref{fig: scatter plots}(h) as scatter plots in order of decreasing coefficient of determination, $R^2$. The best predictors of thermal conductivity are the $[100]$ TA branch sound speed ($R^2=0.89$), the average Gr{\"u}neisen parameter ($R^2=0.85$), and the TA branch $\mathbf{X}$-point frequency ($R^2=0.77$). It is not surprising that these quantities are strong predictors of the thermal conductivity, as the TA branch has a high group velocity and the Gr{\"u}neisen parameter is a measure of anharmonicity. An initially surprising result is that the $[100]$ LA branch sound speed is a poor predictor of thermal conductivity ($R^2=0.23$), as LA phonons, like TA phonons, have high group velocities. The LA sound speed is likely a poor predictor because there is not enough spread in the ensemble predictions to account for the variation in the thermal conductivity ensemble predictions. the COV for the LA sound speed is $0.019$, while that for the TA sound speed is $0.048$.

The worst predictors are the specific heat at a temperature of $300$ K ($R^2=0.00$) and lattice constant (R$^2=0.06$). As with the LA sound speed, both the specific heat ($\sigma=4$ J/kg-K, COV$=0.005$) and lattice constant ($\sigma=0.077\text{ }\A$, COV$=0.014$) have small standard deviations and thus make a minimal contribution to the $24$ W/m-K standard deviation and 0.14 COV of the thermal conductivity ensemble. The low correlation of the lattice constant and thermal conductivity is consistent with our observation that the lattice constant used in the lattice dynamics calculations does not significantly impact the behavior of the thermal conductivity ensemble.

It is instructive to compare Figs.~\ref{fig: scatter plots}(c) and \ref{fig: scatter plots}(e), which plot the thermal conductivity versus the frequency of the TA and LO branches at the $\mathbf{X}$-point. These two dispersion branches have large frequency spreads at the $\mathbf{X}$-point. The TA branch has a standard deviation of $0.57$ THz and that of the LO branch is $0.39$ THz. The TA frequency at the $\mathbf{X}$-point, however, is a better predictor of the thermal conductivity ($R^2=0.77$) than the LO $\mathbf{X}$-point frequency ($R^2=0.42$). Because the LO group velocities are small, they do not make a significant contribution to the thermal conductivity, so that it is not surprising that there is a weak correlation between these quantities.

A quantity that we anticipated would be correlated with thermal conductivity is the three-phonon phase space, which is defined in Sec.~S4. Although the three-phonon phase space is purely a harmonic property, Lindsay and Broido showed that it is inversely correlated to the thermal conductivity of several semiconductors including silicon, \cite{lindsay2} indicating that materials with fewer available scattering processes tend to have higher thermal conductivities. As shown in Fig.~\ref{fig: scatter plots}(d), there is an inverse correlation between the ensemble phase space and thermal conductivity for silicon, but the qualitative behavior is different from that observed by Lindsay and Broido. There is only an inverse correlation for thermal conductivity predictions above $150$ W/m-K ($R^2=0.65$), while the correlation is weak for predictions below $150$ W/m-K ($R^2=0.02$). The five self-consistent XC functionals follow this relationship, with two functionals (LDA and PBEsol) lying in the lower range and the other three (BEEF-vdW, optPBE-vdW, and PBE) lying in the upper range.

\section{Conclusion}

We presented a computationally-efficient framework that uses the BEEF-vdW ensemble to quantify the uncertainty due to XC functional choice in predictions of phonon properties and lattice thermal conductivity. We applied this framework to isotopically-pure silicon, a popular benchmark of \textit{ab initio} predictions of thermal conductivity. As summarized in Table \ref{tbl: si results}, we found that the BEEF-vdW best-fit value bounds most of the self-consistent predictions to within two ensemble standard deviations. This agreement encompasses harmonic quantities such as the phonon frequencies [Figs.~\ref{fig: dispersions}(a) and \ref{fig: dispersions}(b)], specific heat (Fig.~\ref{fig: cp}) and the three-phonon phase space, as well as properties that require incorporating anharmonic effects like the thermal conductivity (Fig.~\ref{fig: ens_k}) and the average Gr{\"u}neisen parameter.

In addition to quantifying the XC uncertainty, our results provide insight into the way that DFT at the GGA-level describes the phonon dynamics in silicon. We found, for example, that the greatest spread in ensemble dispersions occurs at the $\mathbf{X}$-point in the TA branch, in agreement with previous works that found accurate prediction of phonon frequencies in that part of the Brillouin zone to be challenging. We found that ensemble functionals vary widely in their descriptions of phonons with mean free paths between $100$ and $300$ nm, and that these variations are correlated with predictions of thermal conductivity (Fig.~\ref{fig: ens_accum}). As shown in Fig.~\ref{fig: scatter plots}, we were able to use the ensemble to identify the [100] TA sound speed and the average Gr{\"u}neisen parameter as good predictors of thermal conductivity. Conversely, we found that the specific heat and [100] LA sound speed, which are described consistently amongst the ensemble members, are poor predictors of the thermal conductivity despite being essential components of the calculation. Because our framework can be used to examine the predictions of thousands of XC functionals, it can be applied in the future to identify trends in phonon dynamics that occur due to, or in spite of, the choice of XC functional in other materials.

\begin{acknowledgments}
 We thank Ankit Jain, Sushant Kumar, Ravishankar Sundararaman, Gregory Houchins, and Dilip Krishnamurthy for helpful discussions. H. L. P. and V. V. acknowledge support from the Office of Naval Research under Award No. N00014-19-1-2172 and a Presidential Fellowship at Carnegie Mellon University. Acknowledgment is also made to the Extreme Science and Engineering Discovery Environment (XSEDE) for providing computational resources through Award No. TG-CTS180061.
\end{acknowledgments}

\newpage

\bibliography{bibliography}

\newpage

\end{document}

\title{Supplemental Material: Uncertainty quantification in first-principles predictions of phonon properties and lattice thermal conductivity}

\author{Holden L. Parks}
\affiliation{Department of Mechanical Engineering, Carnegie Mellon University, Pittsburgh, Pennsylvania 15213, USA}

\author{Hyun-Young Kim}%
\affiliation{Department of Mechanical Engineering, Carnegie Mellon University, Pittsburgh, Pennsylvania 15213, USA}

\author{Venkatasubramanian Viswanathan}%
% \email{venkvis@cmu.edu}
\author{Alan J. H. McGaughey}%
\email{mcgaughey@cmu.edu}
\affiliation{Department of Mechanical Engineering, Carnegie Mellon University, Pittsburgh, Pennsylvania 15213, USA}

\maketitle

\date{\today}

\clearpage

\section{Additional Silicon Ensemble Results}

\subsection{Lattice constants}

\begin{figure}[h]
\begin{tabular}{c}
  (a) \\[-3pt]
  \includegraphics[width=3.4 in]{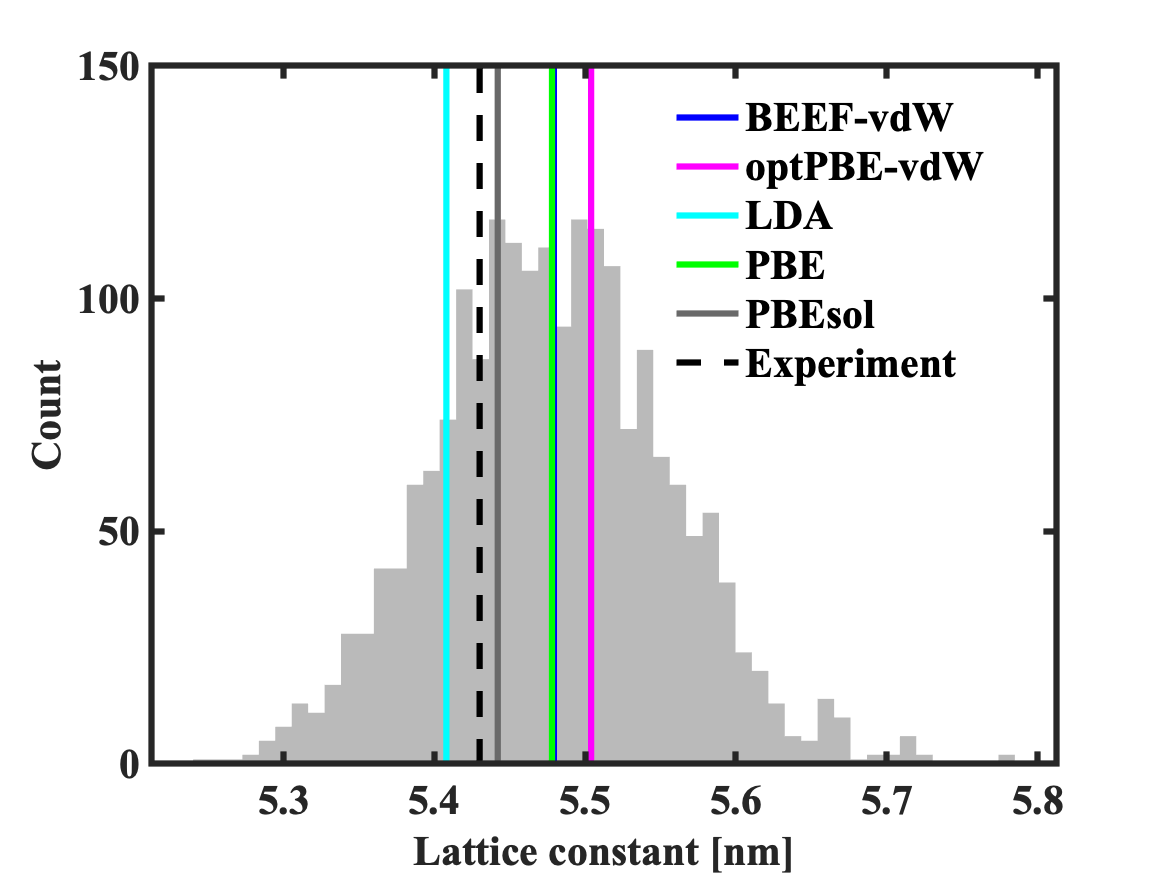}\\
    (b) \\[-3pt]
\includegraphics[width=3.4 in]{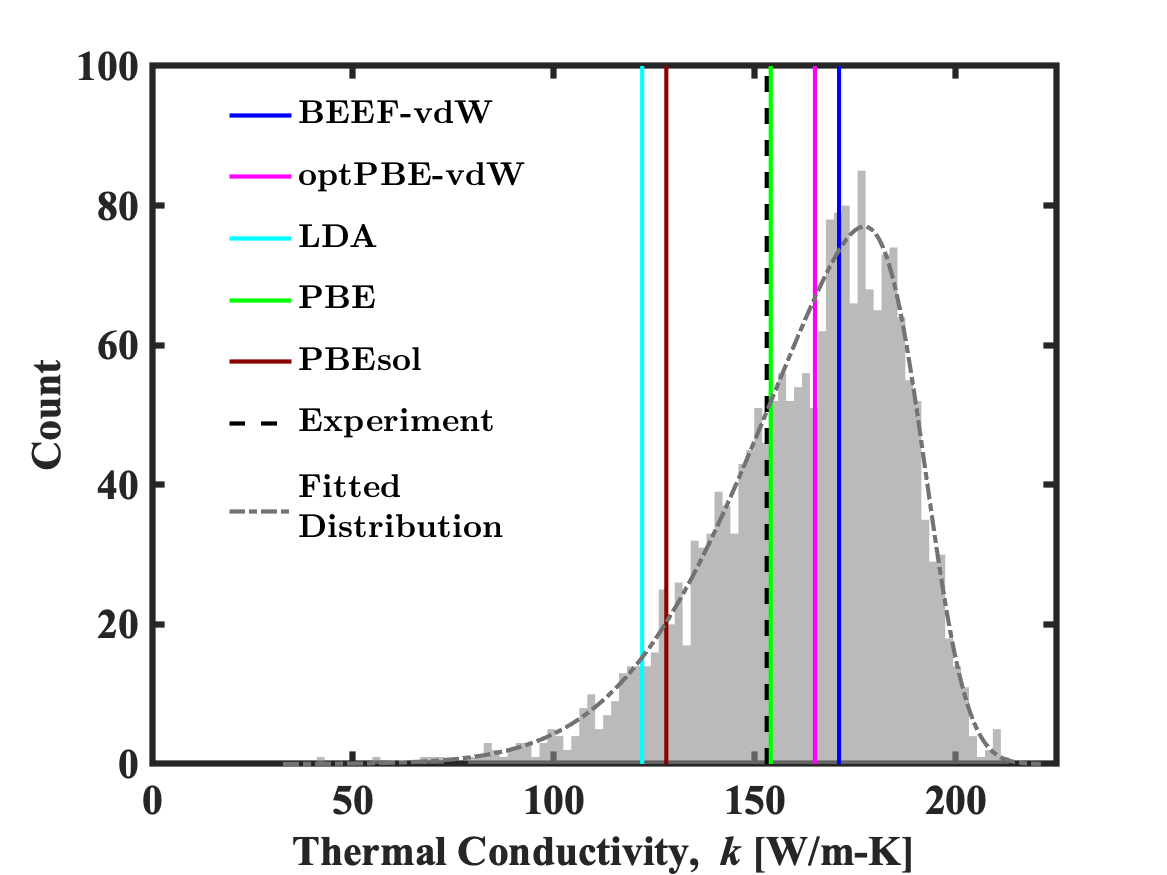}\\
\end{tabular}
\caption{Histograms of (a) ensemble lattice constants for silicon and (b) the ensemble thermal conductivities calculated using the ensemble lattice constants.}
\label{fig: si ens lat results}
\end{figure}

To determine the lattice constant of silicon for a typical DFT calculation, we strain the lattice and fit the resulting energies to the stabilized jellium equation of state. \cite{Alchagirov2003} This fit allows us to determine the lattice constant for which the structure is at its minimum energy. BEEF-vdW yields an ensemble of energy values for every such structure, meaning we can repeat this fitting procedure for each ensemble functional to get an ensemble of lattice constants. See Fig.~1 from Ahmad and Viswanathan \cite{ahmad}, which shows this ensemble of equation of state fits for silicon. The resulting ensemble of lattice constants is plotted in Fig.~\ref{fig: si ens lat results}(a) as a histogram. 

In Sec.~IIIE, we discuss the impact of using the ensemble lattice constants from Fig.~\ref{fig: si ens lat results}(a) in the ensemble lattice dynamics calculations. A histogram of the ensemble predictions of thermal conductivity when the ensemble lattice constants are used, rather than the BEEF-vdW best-fit lattice constant, in shown in Fig.~\ref{fig: si ens lat results}(b).

\newpage

\subsection{Phonon dispersion}

\begin{figure}[h]
\begin{tabular}{ccc}
(a) & (b) & (c) \\[-3pt]
  \includegraphics[width=2.0 in]{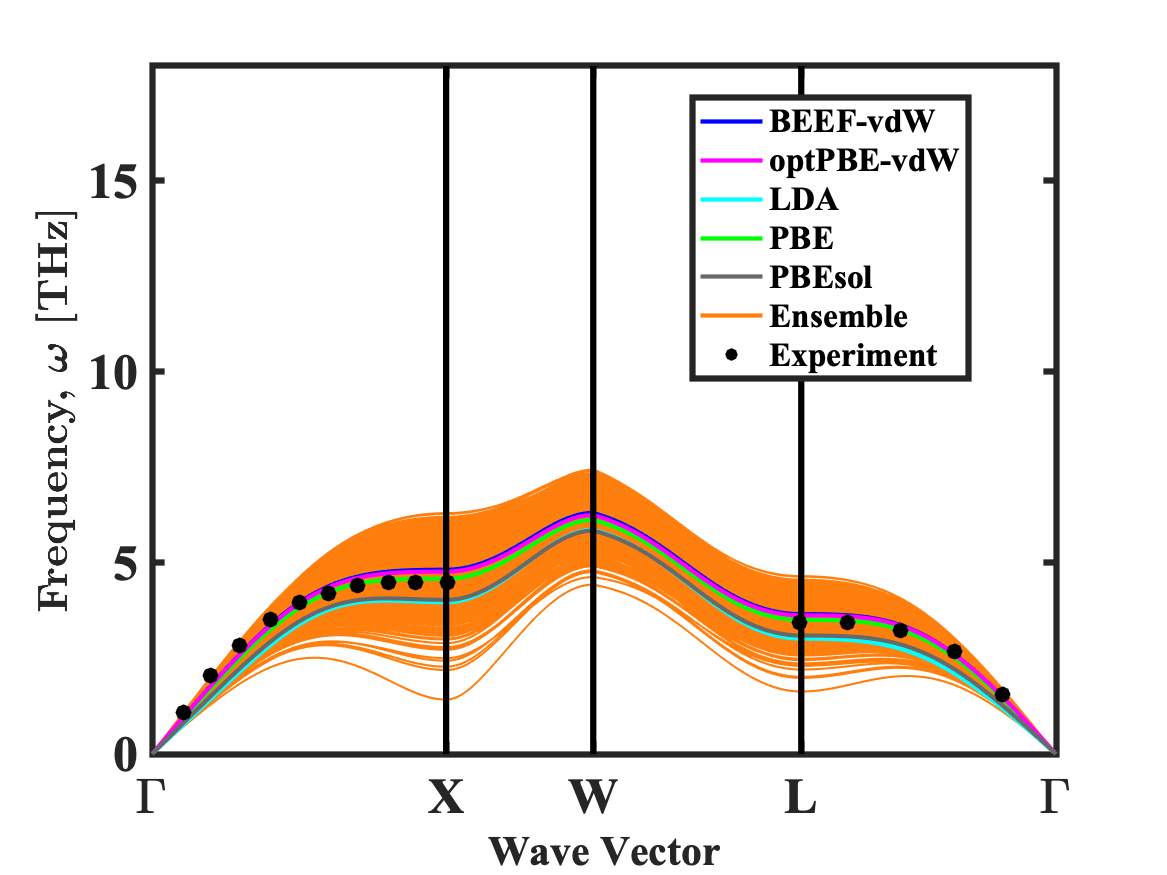} &  \includegraphics[width=2.0 in]{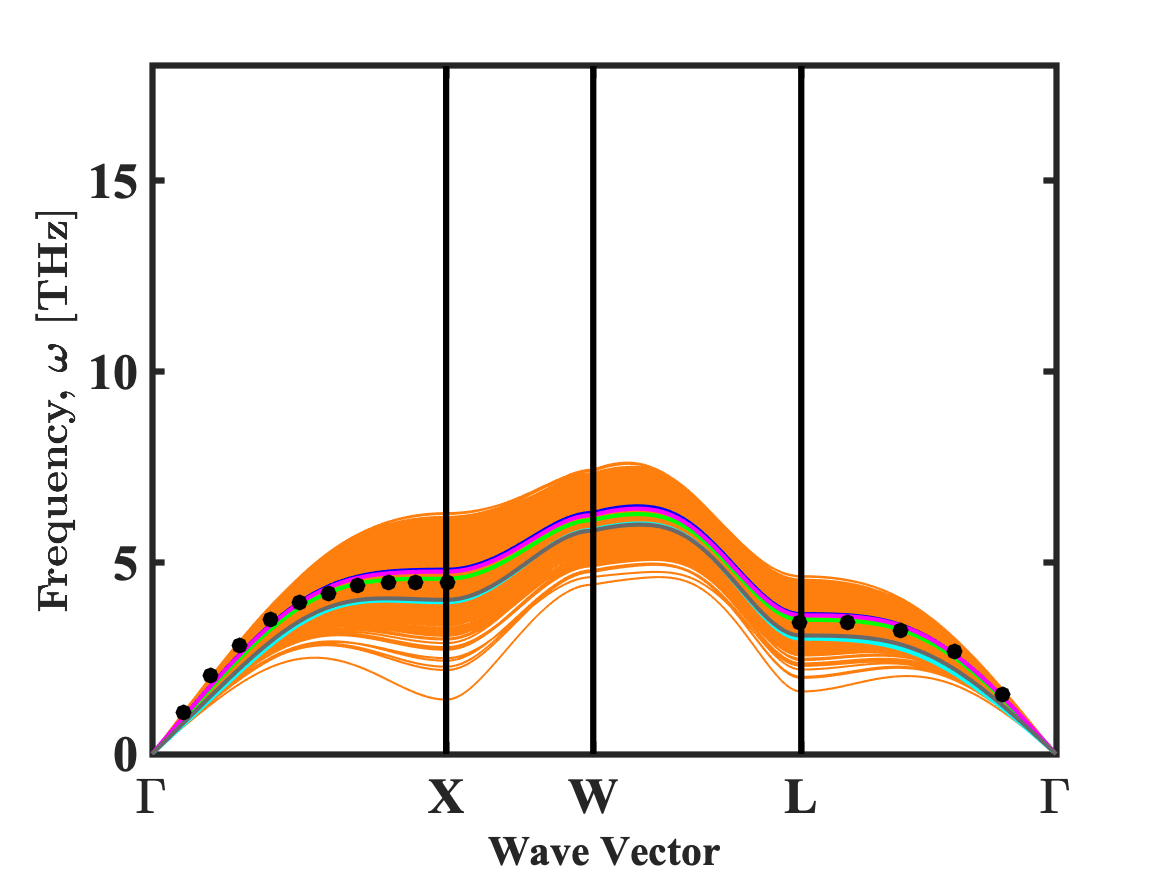} & \includegraphics[width=2.0 in]{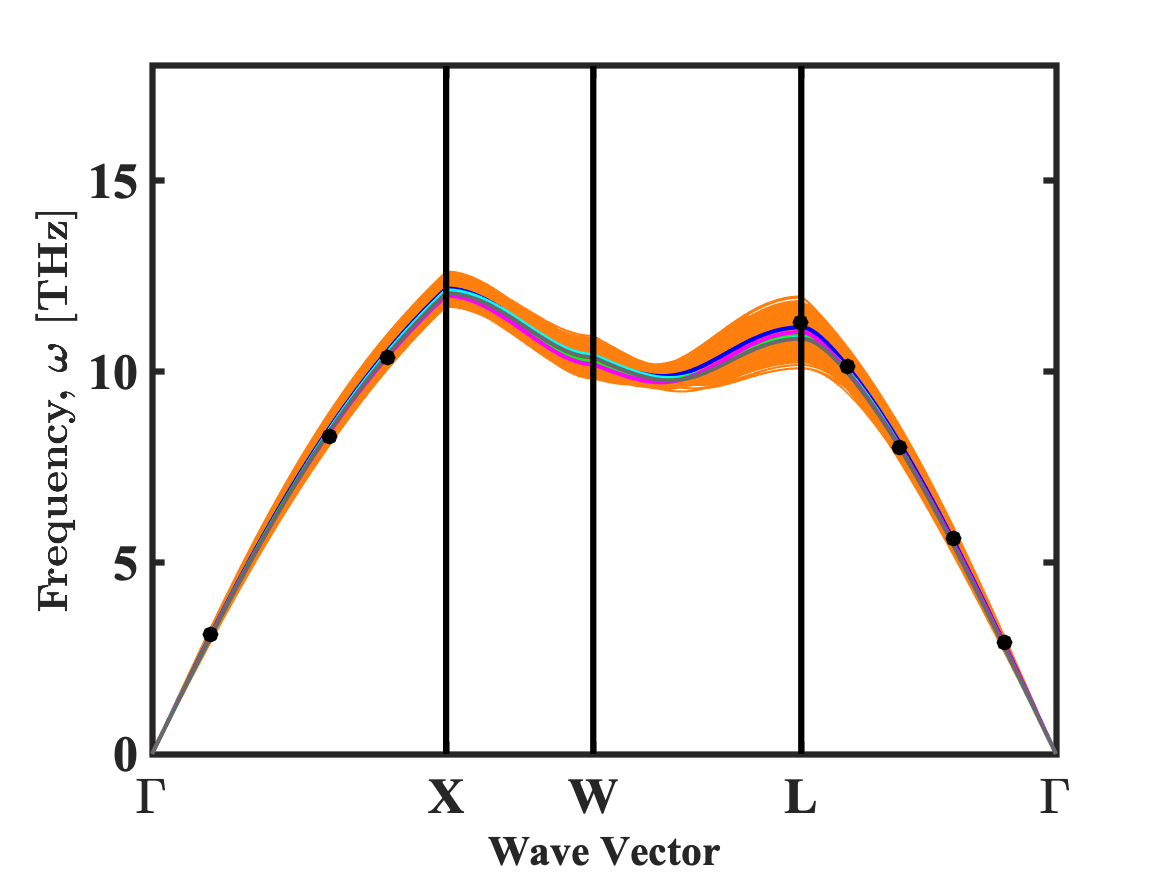}\\
   (d) & (e) & (f)  \\[-3pt] 
 \includegraphics[width=2.0 in]{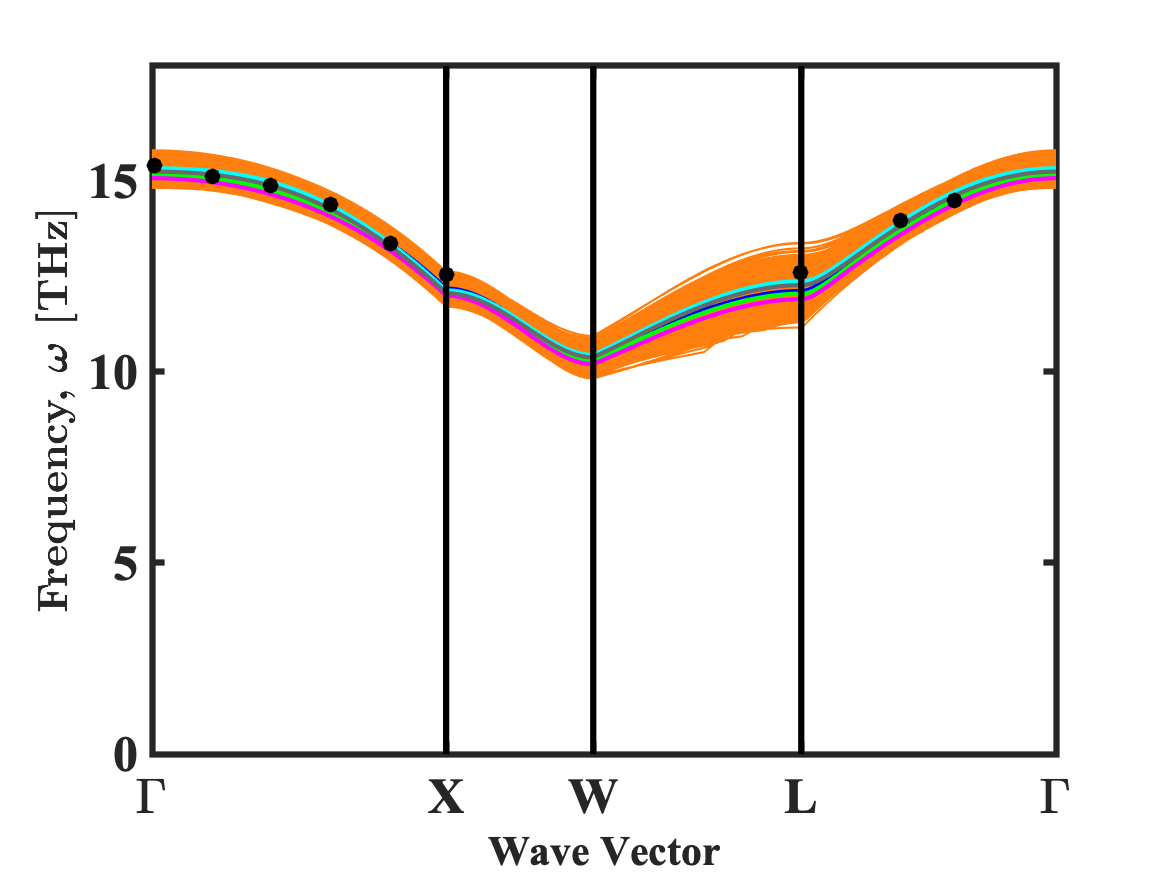} & \includegraphics[width=2.0 in]{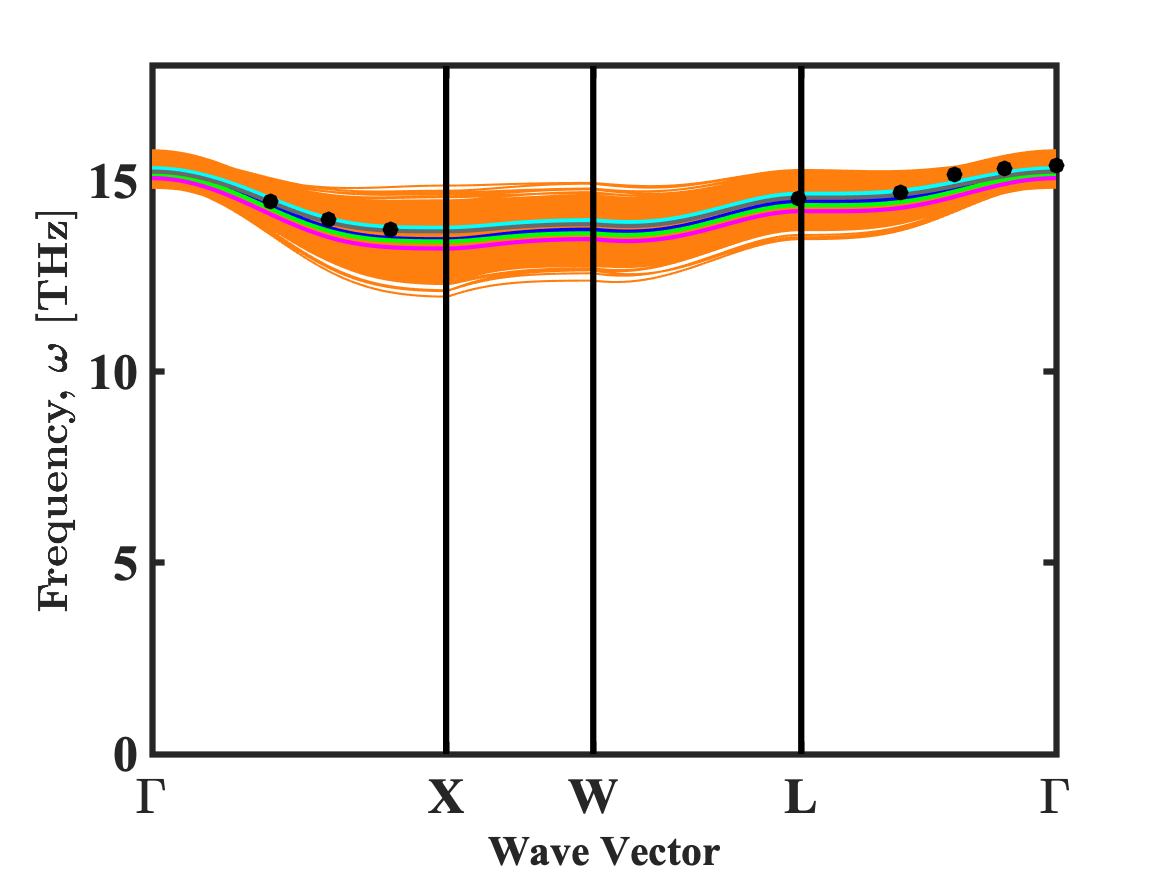} &  \includegraphics[width=2.0 in]{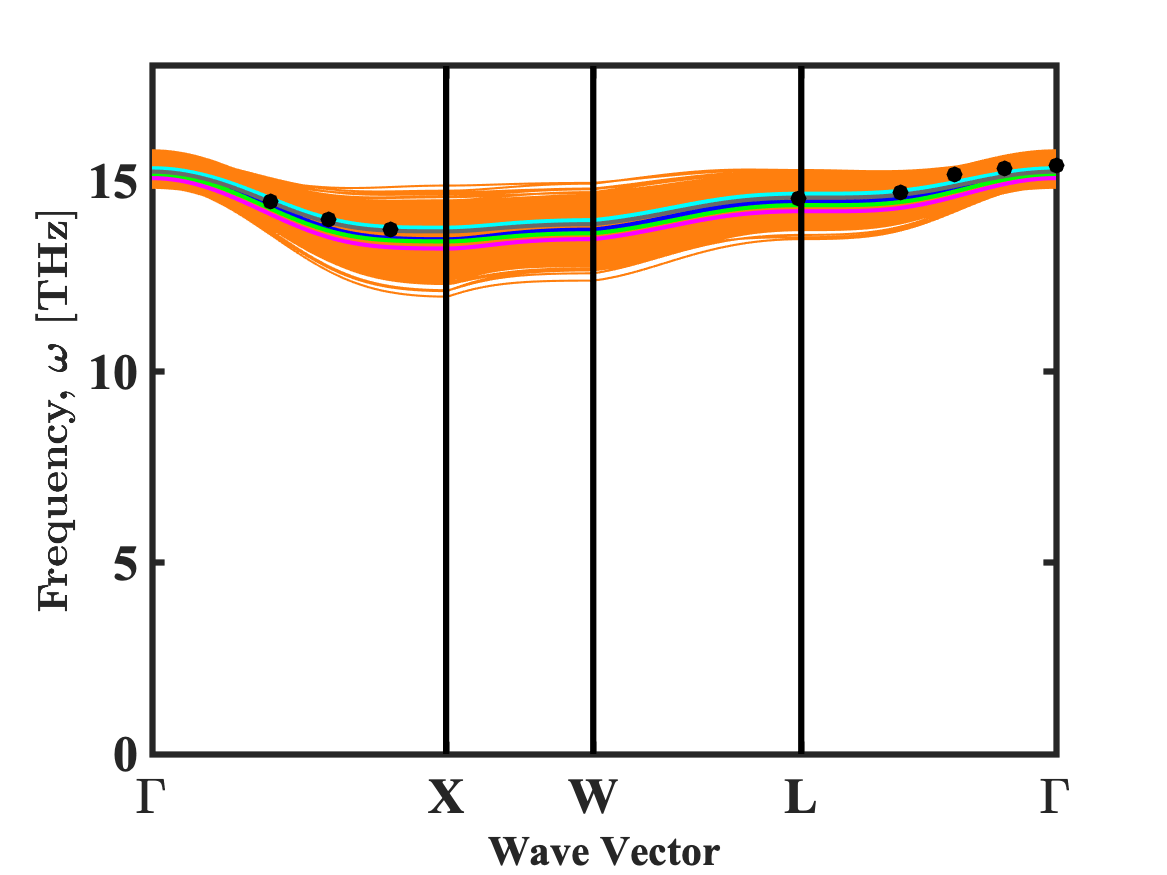}\\
\end{tabular}
\caption{Silicon ensemble phonon dispersions for the (a), (b) TA, (c) LA (d) TO, and (e), (f) LO branches. Experimental values (black dots) are from Nilsson and Nelin. \cite{nilsson}}
\label{fig: si all disp}
\end{figure}

The six phonon dispersion branches of silicon are plotted in Figs.~\ref{fig: si all disp}(a) - \ref{fig: si all disp}(f). The TA [(a) and (b)] and LO [(e) and (f)] branches are degenerate except on the $\mathbf{W}-\mathbf{L}$ path.

\newpage

\subsection{[100] Sound speed histograms}

\begin{figure*}[h]
\includegraphics[width=3.4in]{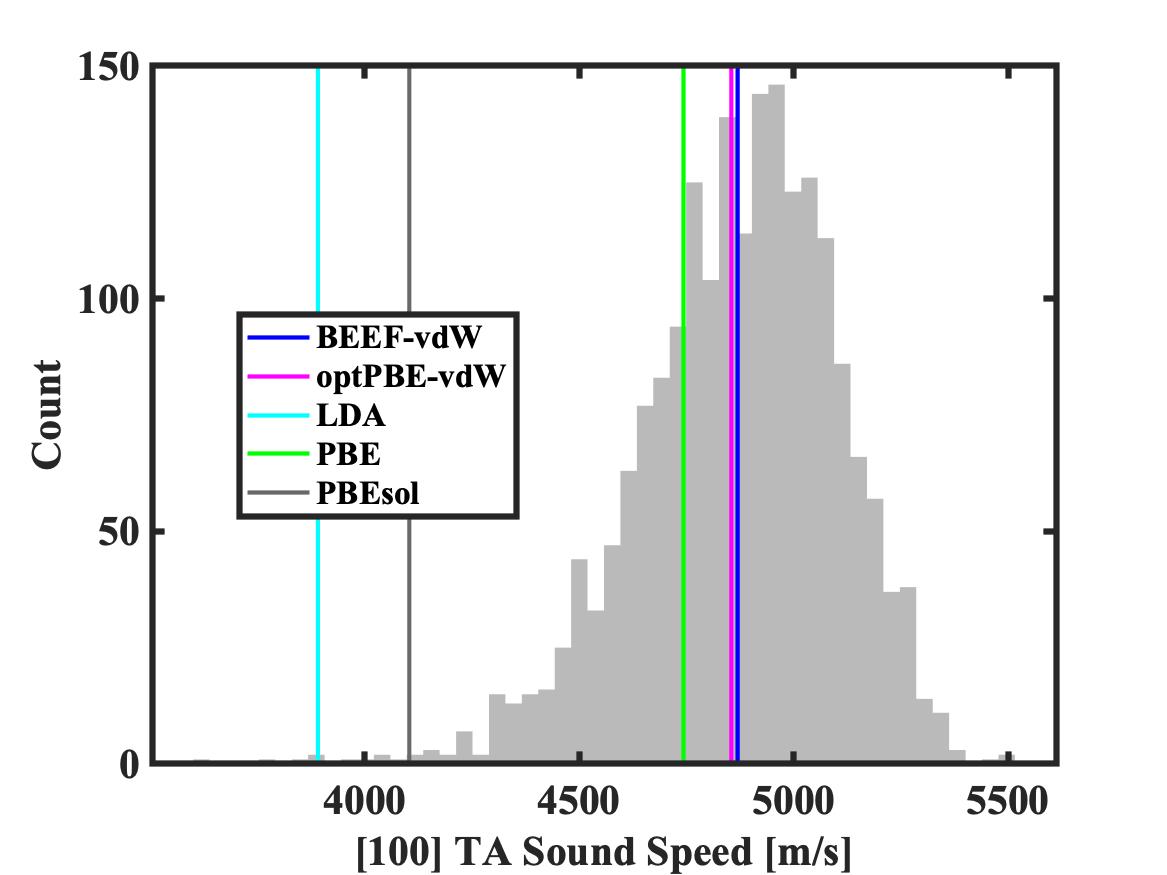}
\caption{Ensemble sound speed of the [100] TA branch.}
\label{fig: ens si ta vs}
\end{figure*}

\begin{figure*}[h]
\includegraphics[width=3.4in]{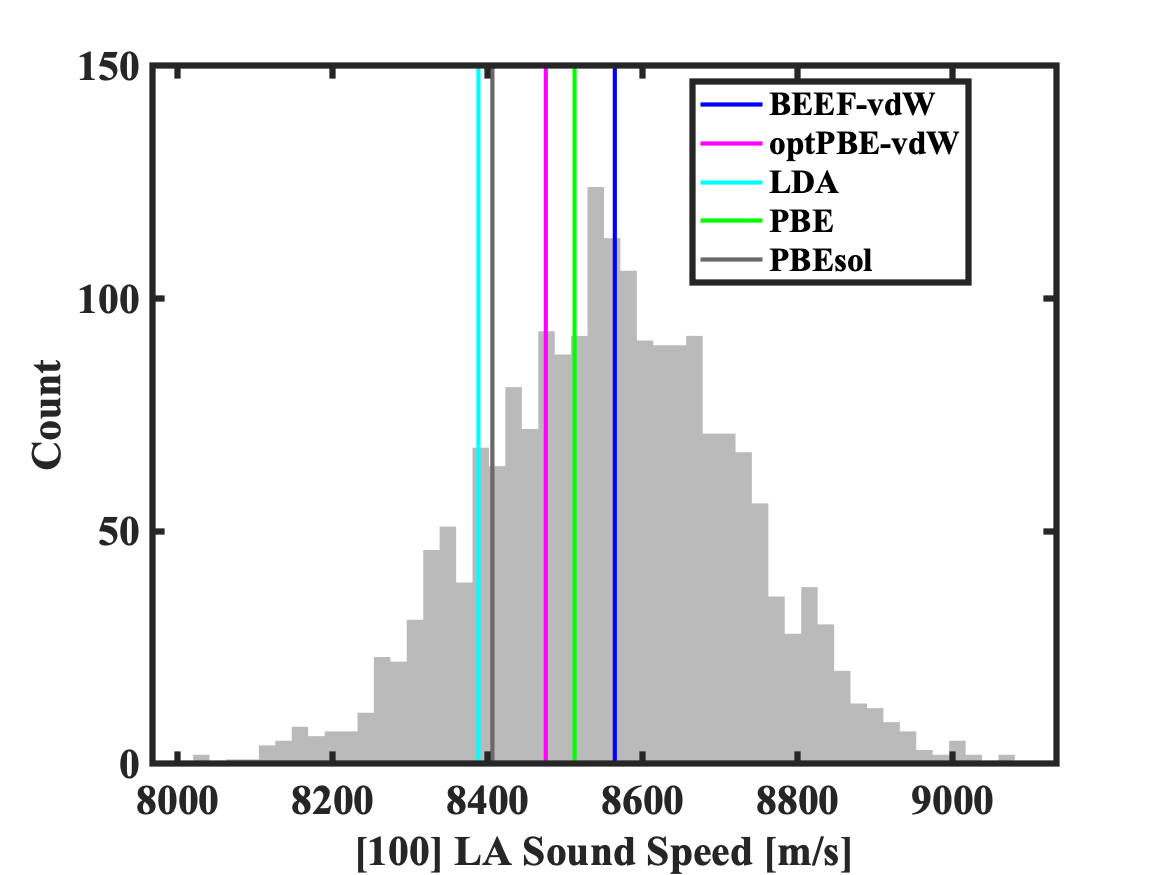}
\caption{Ensemble sound speed of the [100] LA branch.}
\label{fig: ens si la vs}
\end{figure*}

The ensemble histograms of the TA branch $[100]$ sound speed and the LA branch $[100]$ sound speed are plotted in Figs.~\ref{fig: ens si ta vs} and \ref{fig: ens si la vs}.

\newpage

\subsection{Mode-dependent Gr{\"u}neisen parameter}

\begin{figure}[h]
\begin{tabular}{ccc}
(a) & (b) & (c) \\[-3pt]

  \includegraphics[width=2.0 in]{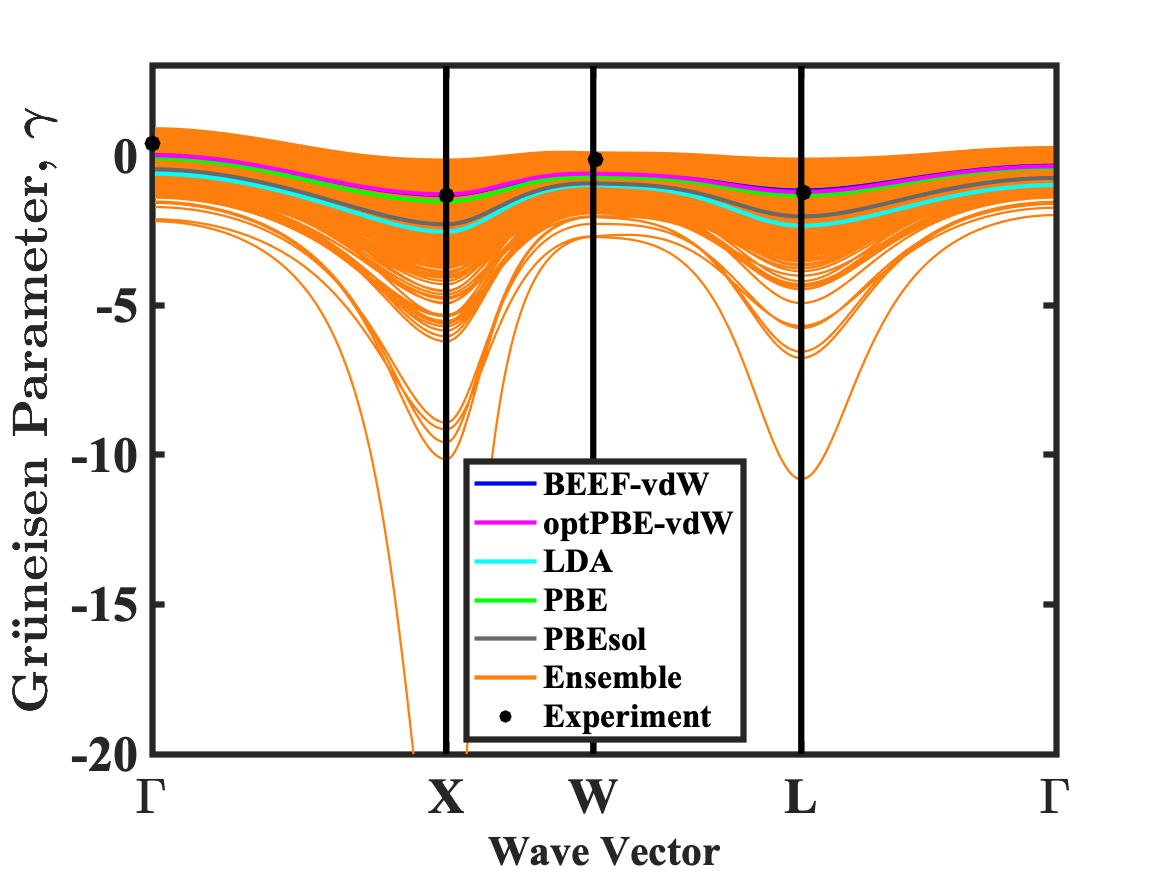} &  \includegraphics[width=2.0 in]{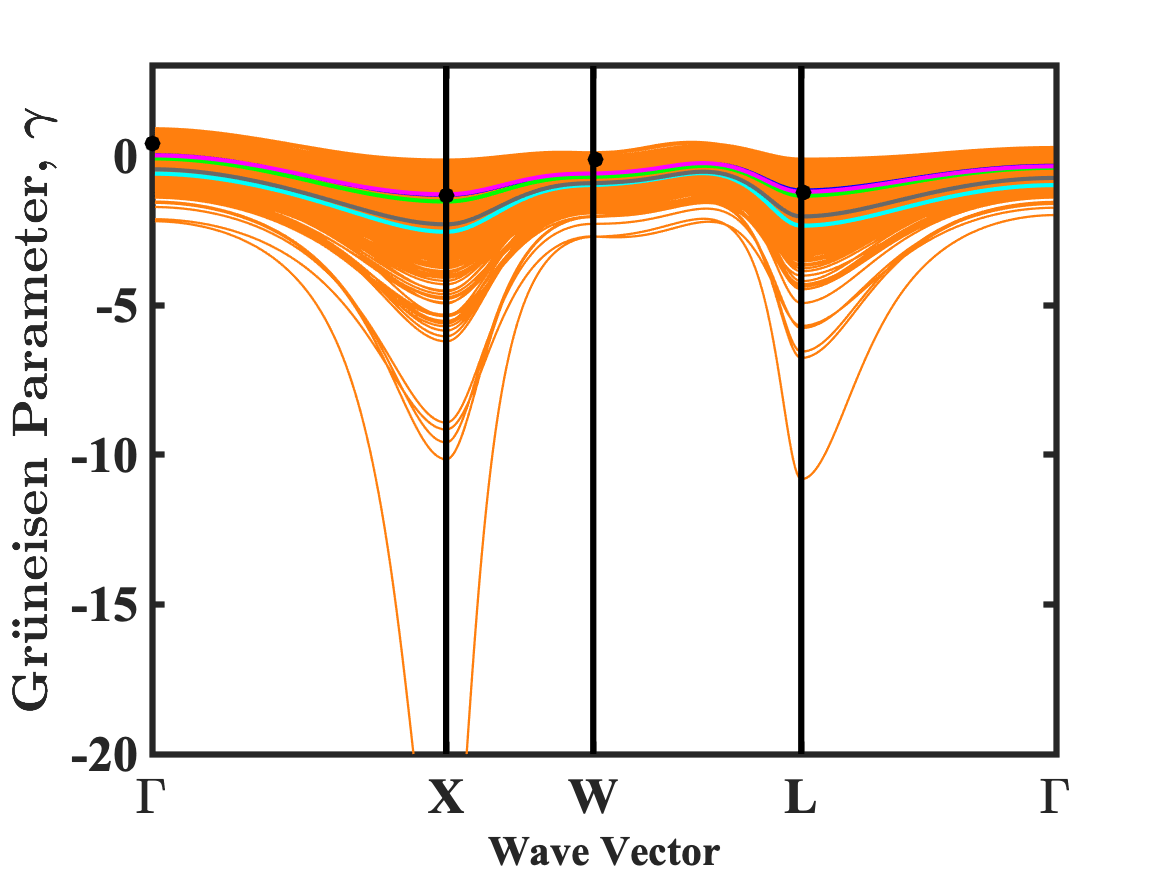} & \includegraphics[width=2.0 in]{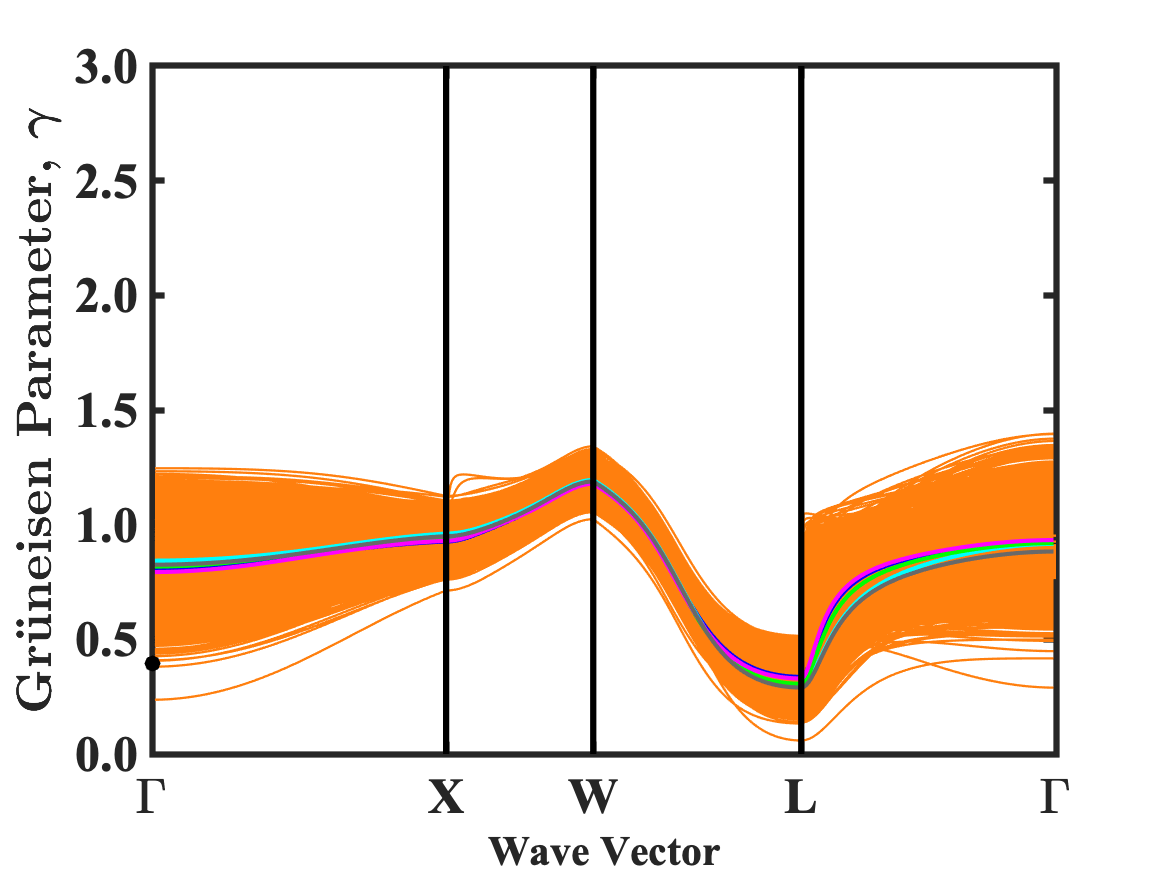}\\
   (d) & (e) & (f)  \\[-3pt] 
 \includegraphics[width=2.0 in]{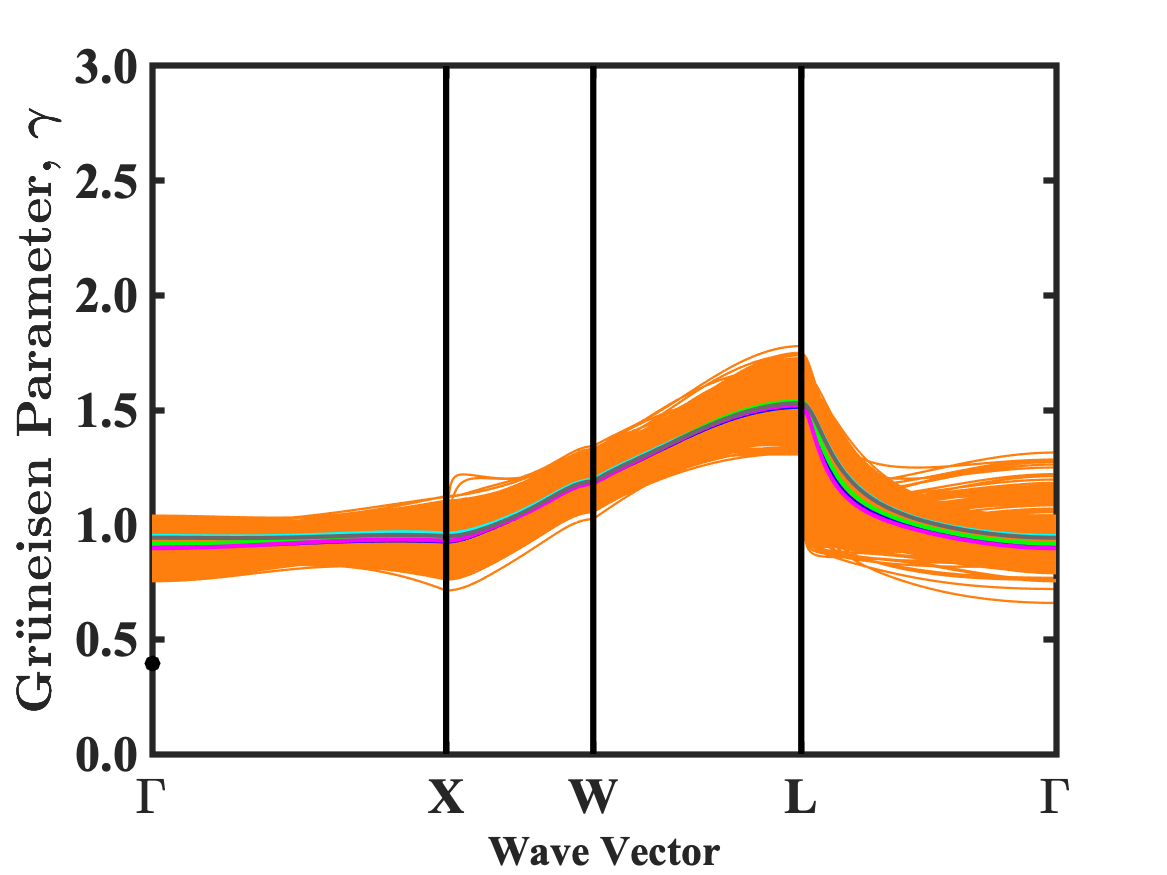}& \includegraphics[width=2.0 in]{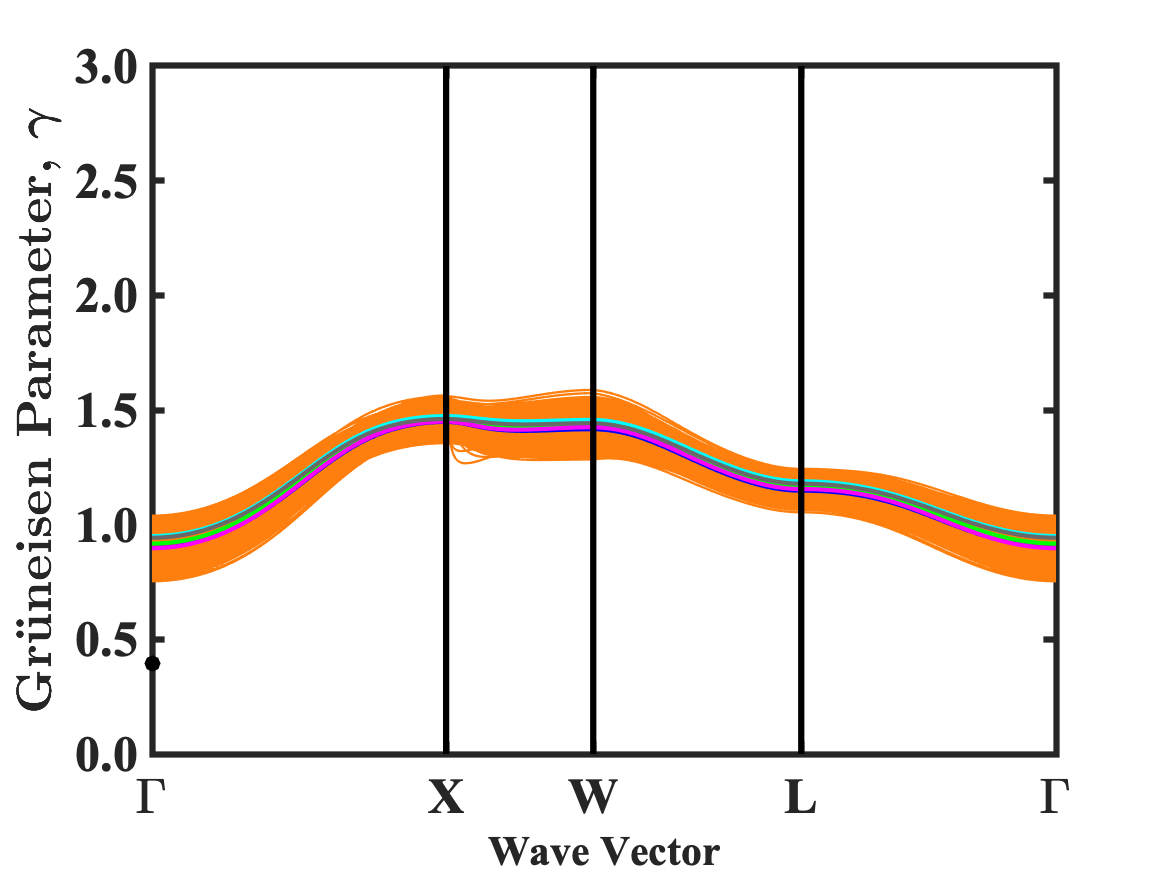} &  \includegraphics[width=2.0 in]{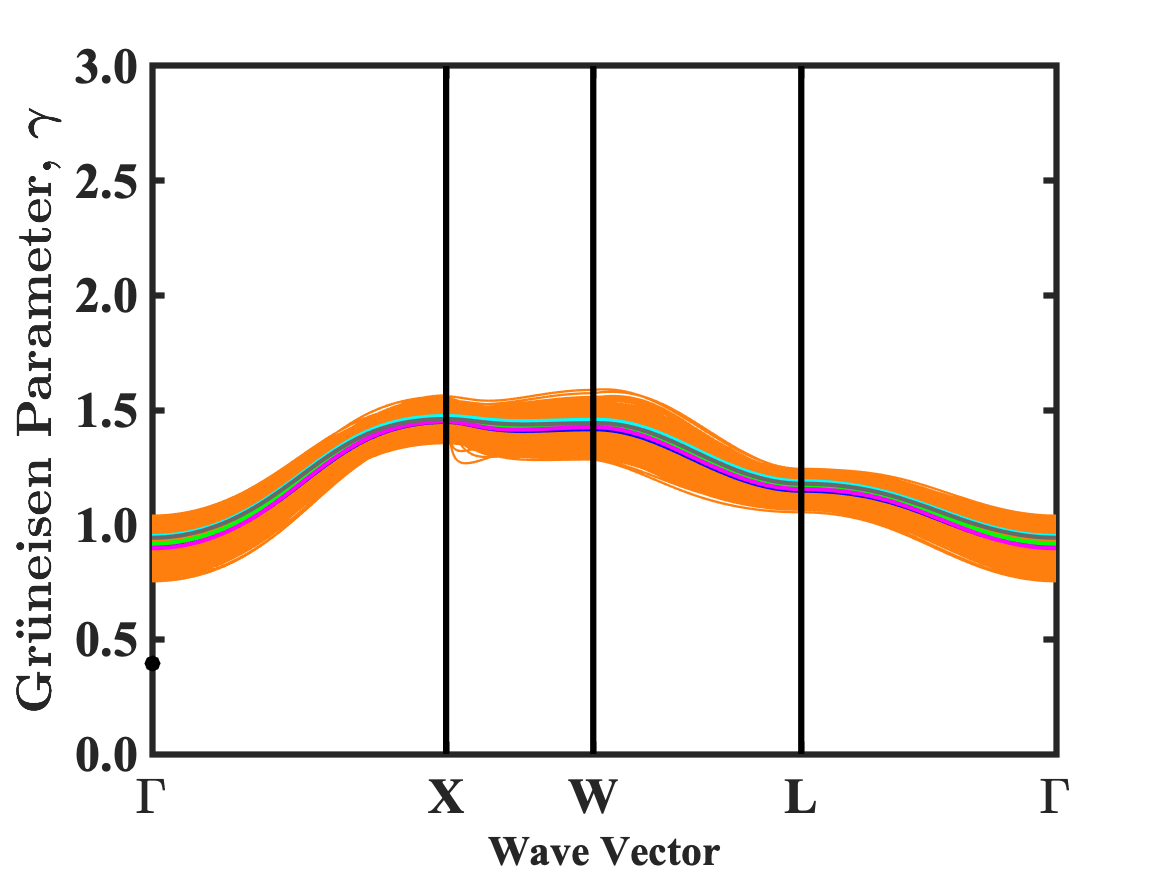}\\
\end{tabular}
\caption{Silicon ensemble Gr{\"u}neisen parameters for the (a), (b) TA, (c) LA, (d) TO, and (e), (f) LO branches. Experimental values (black dots) are from Madelung, R{\"o}ssler, and Schulz. \cite{si_grun}}
\label{fig: si all grun}
\end{figure}

The mode-dependent Gr{\"u}neisen parameters for all six dispersion branches of silicon are plotted in Figs.~\ref{fig: si all grun}(a)-\ref{fig: si all grun}(a). The TA [(a) and (b)] and LO [(e) and (f)] branches are degenerate except on the $\mathbf{W}-\mathbf{L}$ path.

\newpage

\subsection{Distribution fitting}

We applied the workflow outlined by Guan et al. \cite{guan} to determine the model distribution that best describes the BEEF-vdW ensemble data for a variety of thermodynamic quantities for silicon. The ``best'' fit is determined using the Cramer von Mises goodness of fit test. \cite{anderson1952} The model distributions tested include, but are not limited to, the normal distribution and its transformations (e.g., skewed normal, log normal, etc.), Weibull, generalized gamma, Pareto, and chi-squared distributions. The properties for which distributions were fitted are summarized in Table \ref{tbl: si quantities}. The bulk modulus ($B$) and volume ($V$) were calculated by fitting an equation of state and finding the corresponding minimum energy lattice constant as described in Section II C. The bulk modulus is determined from the equation of state through the relation $B=V(\partial^2 U/\partial V^2 )_T$ \cite{kittel}, where $U$ is the potential energy of the system and the $T$ subscript indicates that temperature is held constant. This derivative can be determined analytically from the equation of state. The mean Gr{\"u}neisen parameter is calculated as \cite{ritz}

\begin{equation}
\bar{\gamma} = \frac{\sum_{\mathbf{q},\nu} c(\mathbf{q}, \nu)\gamma(\mathbf{q},\nu)}{\sum_{\mathbf{q},\nu} c(\mathbf{q},\nu)}.
\end{equation}

\noindent Note that we do not take the absolute value of $\gamma(\mathbf{q},\nu)$ as we do in the main text, so as to be consistent with Ritz et al.\cite{ritz} The thermal expansion coefficient (TEC) $\alpha$ can be calculated as
\begin{equation}
\alpha = \frac{\bar{\gamma} c}{3B}.
\end{equation}

\noindent where $c$ is the volumetric specific heat.

\begin{figure}[h]
\begin{tabular}{ccc}
(a) & (b) & (c) \\[-3pt]
  \includegraphics[width=2.0 in]{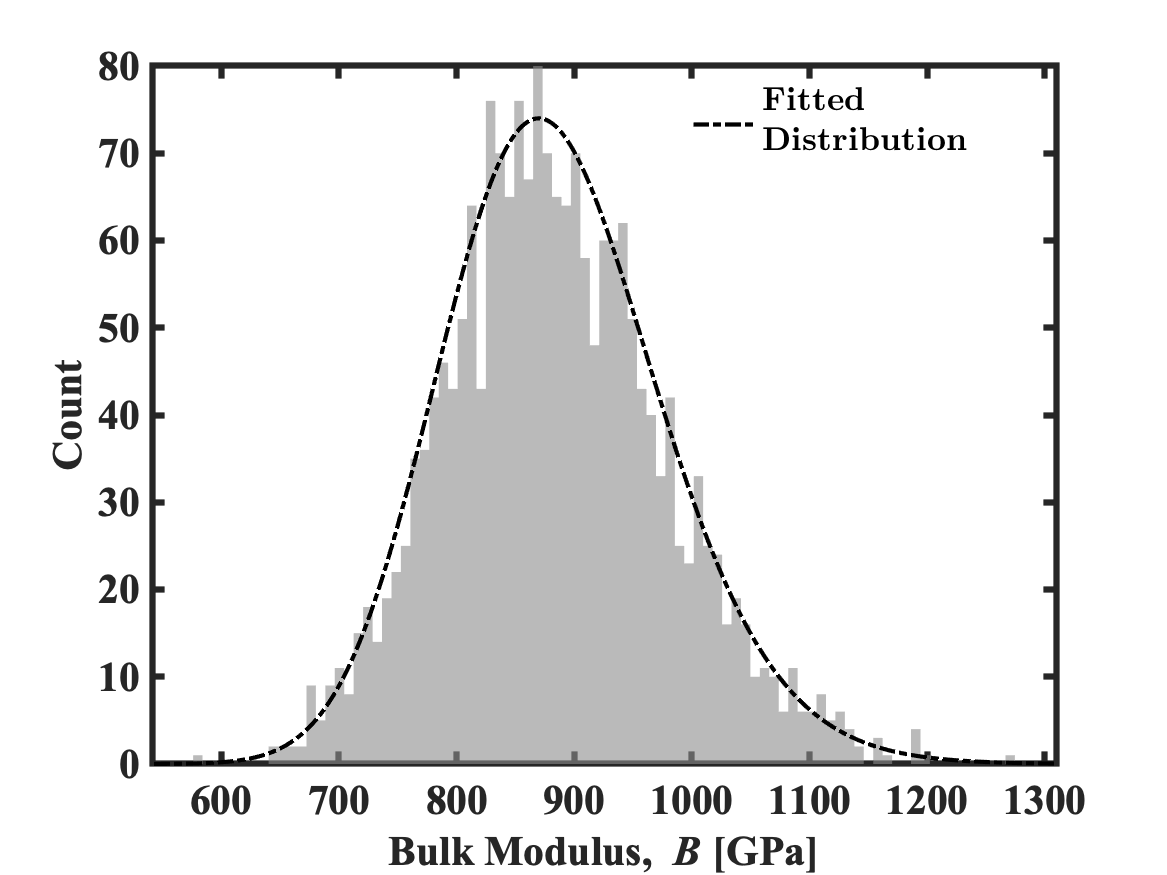} &  \includegraphics[width=2.0 in]{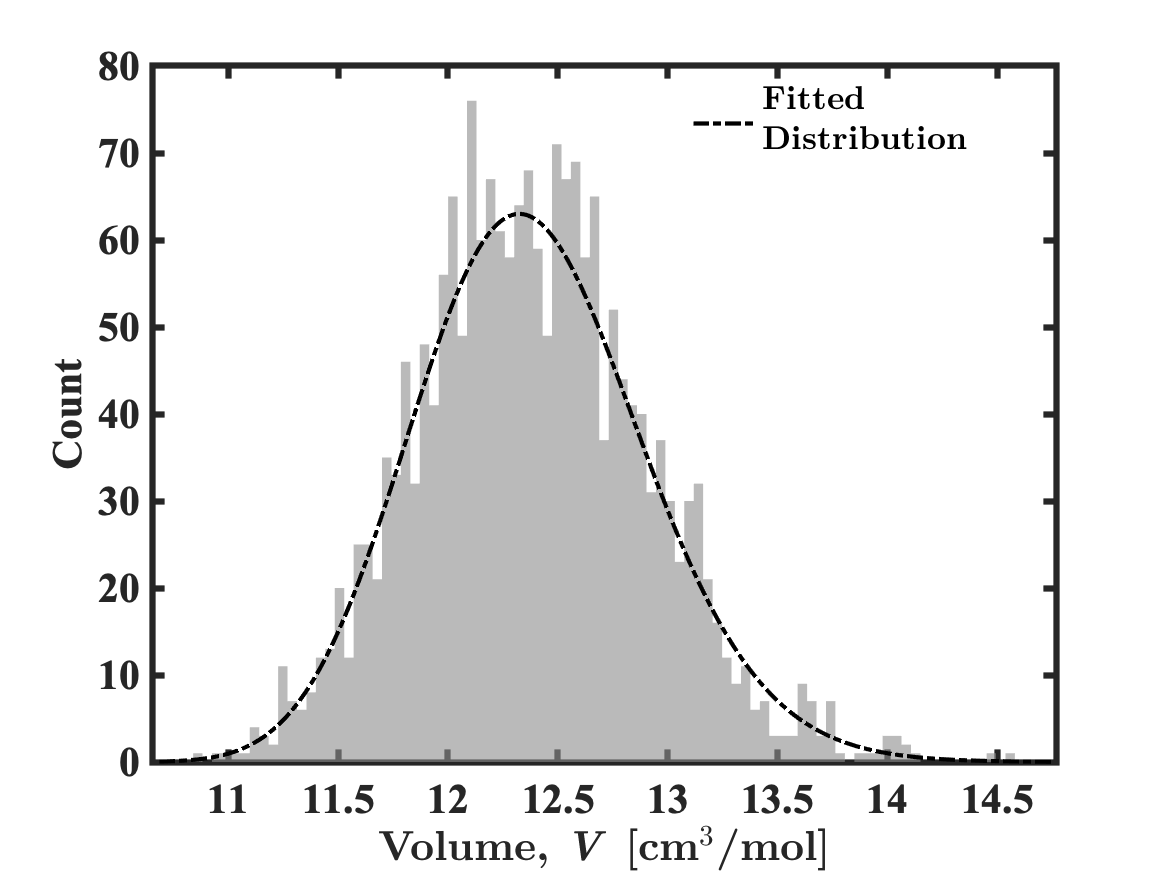} & \includegraphics[width=2.0 in]{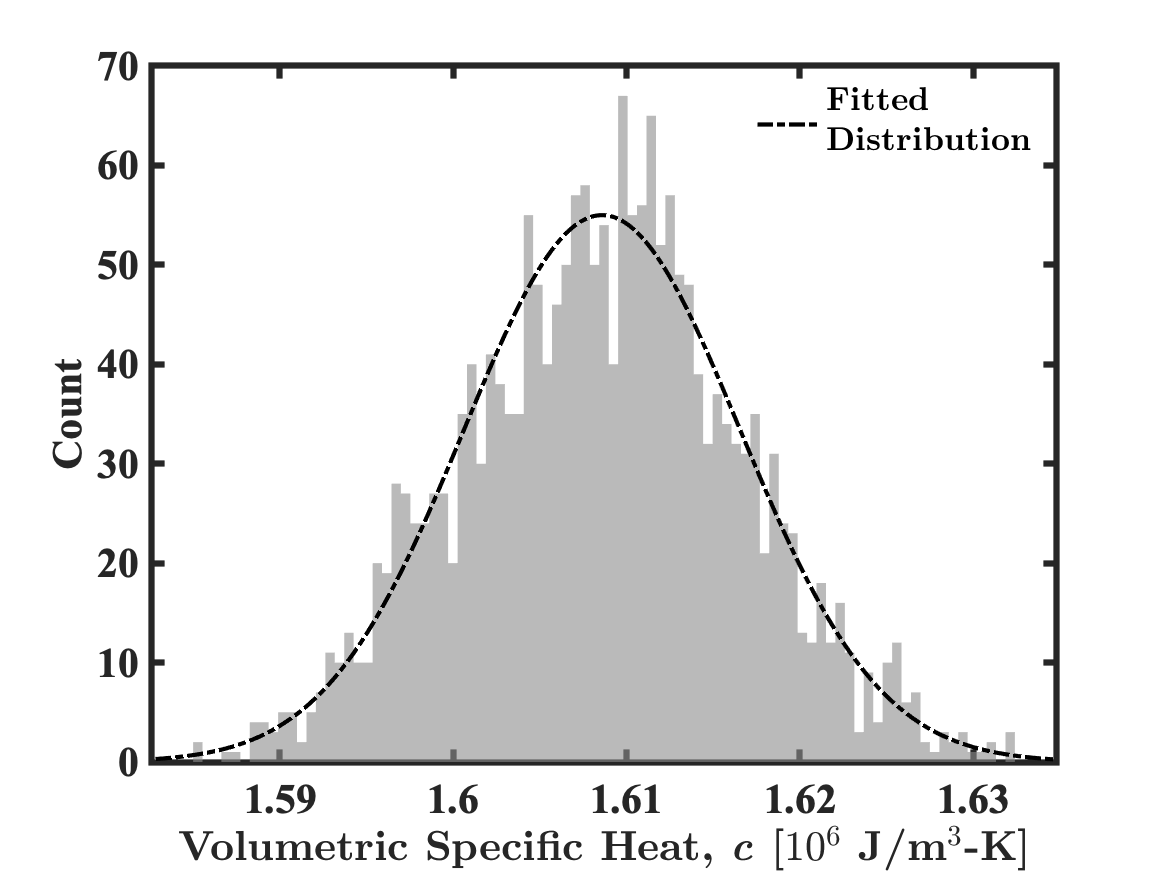}\\
   (d) & (e) & (f)  \\[-3pt] 
 \includegraphics[width=2.0 in]{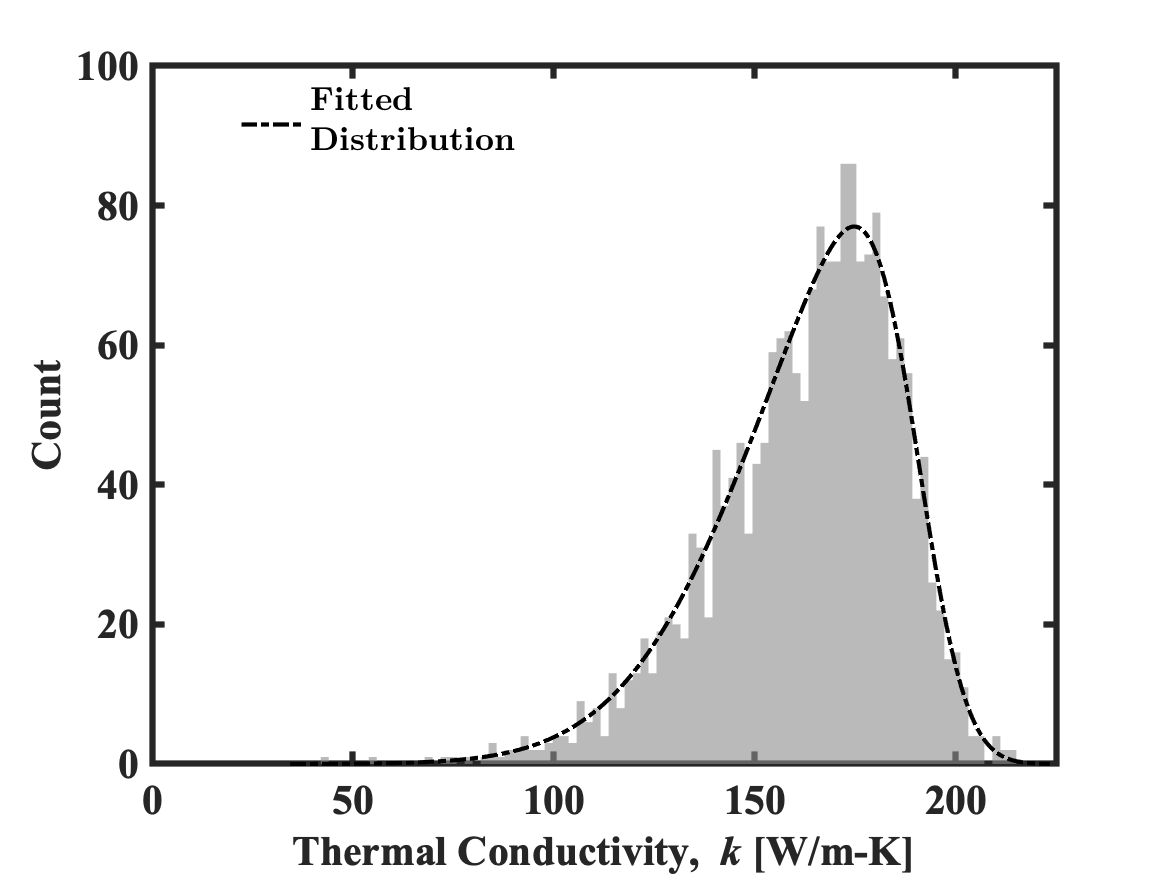}& \includegraphics[width=2.0 in]{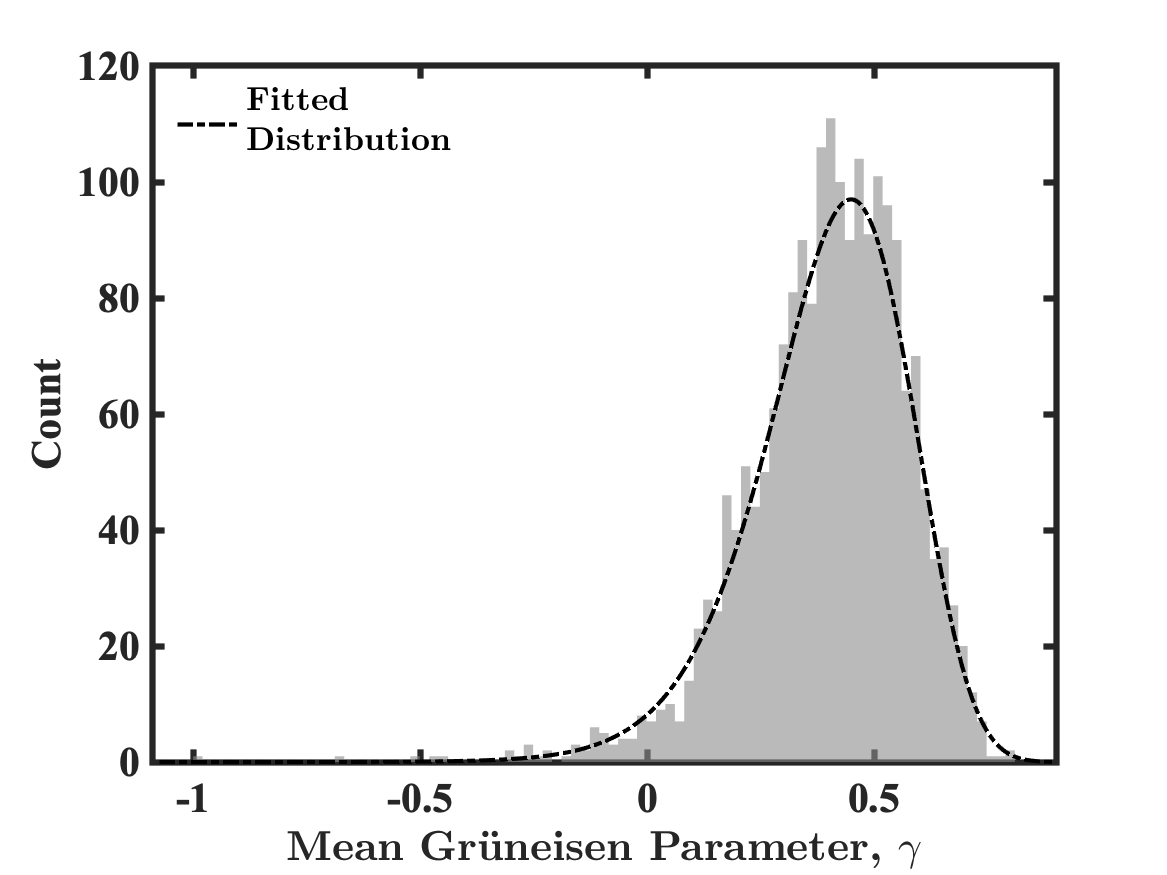} &  \includegraphics[width=2.0 in]{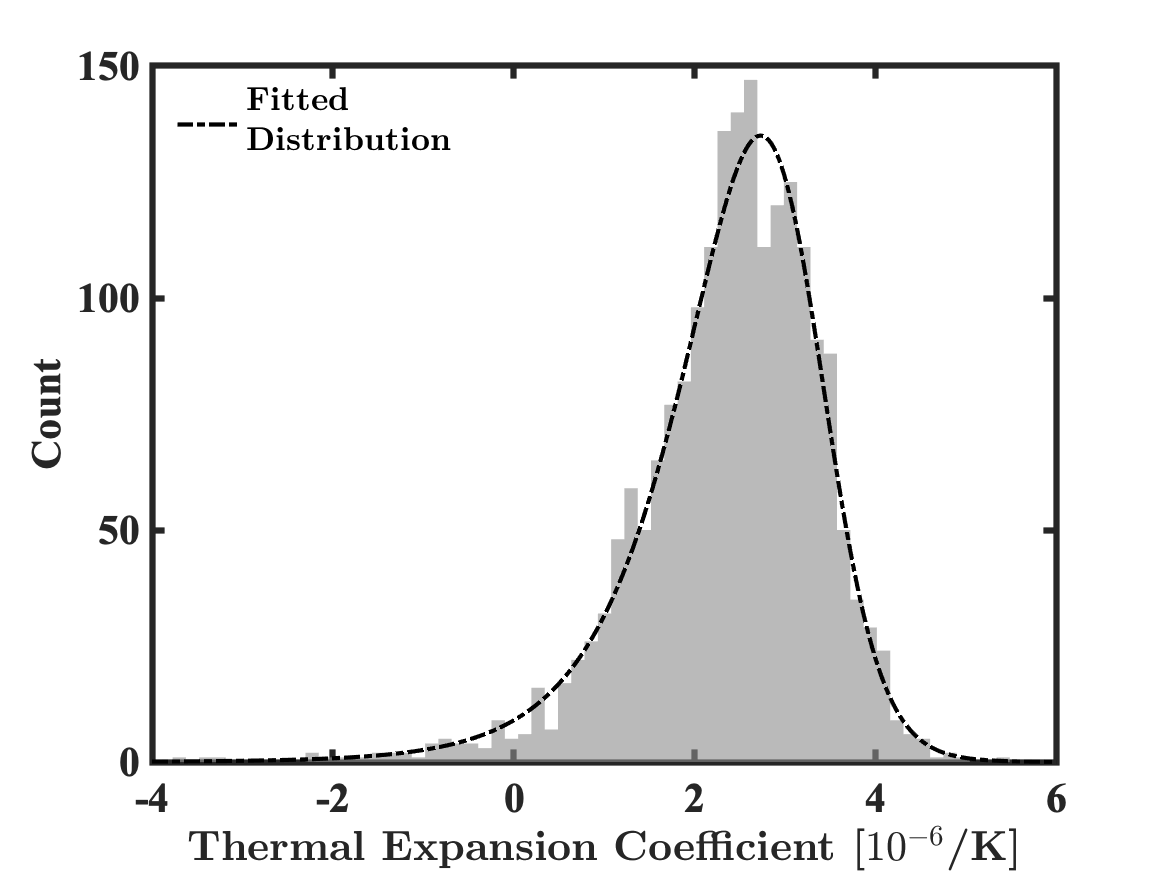}\\
\end{tabular}
\caption{BEEF-vdW ensembles for the quantities summarized in Table~\ref{tbl: si quantities}. The overlaid distributions are fitted using the Cramer von Mises goodness of fit test as described by Guan et al. \cite{guan} and are described in Table~\ref{tbl: si distributions}.}
\label{fig: dist fit}
\end{figure}

The distributions for each quantity are summarized in Table \ref{tbl: si distributions} and are plotted with their respective BEEF-vdW ensembles in Fig.~\ref{fig: dist fit}. Compared to the TEC distributions provided by Guan et al. (who did not model silicon), our silicon TEC ensemble distribution has a higher negative skewness (-1.48) but is similar in terms of its high excess kurtosis (6.72 for silicon, compared to 4.12 for the GaAs ensemble from Guan et al.). Silicon has a lower TEC compared to any of the materials tested by Guan et al., so we use the coefficient of variation (COV), $\sigma/\mu$, as a metric to compare the relative spread of the distributions.\cite{beef_molecules,guan}. The silicon TEC distribution has a COV of $0.40$, which is comparable to the COV of most TEC ensembles at $T=300$ K from Guan et al.

\begingroup
\squeezetable
\begin{table}[h]
\caption{\label{tbl: si quantities}Predicted bulk modulus, molar volume, volumetric specific heat (at $T=300$ K), thermal conductivity (at $T=300$ K), thermal expansion coefficient ($\alpha$), and mean Gr{\"u}neisen parameter ($\bar{\gamma}$) of silicon using different XC functionals. BEEF-vdW predictions include a ``$\pm\sigma$'' term, where $\sigma$ is the standard deviation of the BEEF-vdW ensemble for that quantity.} 
\begin{ruledtabular}
\begin{tabular}{>{\RaggedLeft\arraybackslash}p{2cm}|>{\RaggedLeft\arraybackslash}p{2cm}|>{\RaggedLeft\arraybackslash}p{2cm}|>{\RaggedLeft\arraybackslash}p{2.4cm}|>{\RaggedLeft\arraybackslash}p{2cm}|>{\RaggedLeft\arraybackslash}p{2cm}|>{\RaggedLeft\arraybackslash}p{2.4cm}}
 XC Functional & $B$ (GPa) & $V$ (cm$^3$/mol) & $C_{V}$ ($10^6$J/m$^3$-K) & $k$ (W/m-K) &  $\bar{\gamma}$ & $\alpha$ ($10^{-6}$ K$^{-1}$ ) \\
\hline
BEEF-vdW & $87.2\pm 9.4$ & $12.38\pm0.52$ & $1.61\pm 0.01$ & $171\pm 24$ & $0.42\pm0.18$ & $2.60\pm1.04$  \\
optPBE-vdW & 83.9 & 12.55 & 1.60 & 165 & 0.43 & 2.71  \\
LDA & 94.8 & 11.91 & 1.68 & 122 & 0.20 &  1.18  \\
PBE & 86.9 & 12.38 & 1.62 & 154 & 0.39 & 2.39  \\
PBEsol & 91.8 & 12.13 & 1.61 & 128 & 0.25 & 1.45  \\
\end{tabular}
\end{ruledtabular}
\end{table}
\endgroup

\begingroup
\squeezetable
\begin{table}[h]
\caption{\label{tbl: si distributions}Best-fit distributions determined using the Cramer von Mises test for the BEEF-vdW ensembles of bulk modulus ($B$), molar volume ($V$), specific heat at $T=300$ K (here, $c_{ph}$), thermal conductivity at $T=300$ K ($k$), thermal expansion coefficient ($\alpha$), and average Gr{\"u}neisen parameter ($\bar{\gamma}$) for silicon. ``Form'' refers to the equation for the probability density function for the random variable $x$ (or, in the case of $\alpha$ and $\bar{\gamma}$, for the shifted random variable $y$). $\phi$ and $\Phi$ refer to the probability distribution function and cumulative distribution function of the standard normal distribution. The Cramer von Mises statistic (CVM) for the fit is presented in the final column. Shape parameters (i.e., $a$, $b$, $c$, and $s$) do not have units, while $\mu$ and $\sigma$ have units of the described quantity (e.g., they have units of GPa for the distribution of $B$).}
\begin{ruledtabular}
\renewcommand{\arraystretch}{1.5}
\begin{tabular}{ccccc}
 Quantity & Distribution & Form & Parameters & CVM   \\
\hline
$B$ & Johnson SU & $\frac{b}{\sqrt{x^2+1}}\phi\left(a+b\log(x+\sqrt{x^2+1}); \mu, \sigma\right)$ & $a=-5.28$, $b=6.39$, $\mu=47.80$, $\sigma=43.48$ & 0.99 \\
$V$  & Power Log-Normal & $\frac{c}{x\cdot s}\cdot\phi\left(\frac{\log(x)}{s};\mu, \sigma \right)\cdot \Phi\left(-\frac{\log{x}}{s};\mu, \sigma \right)^{c-1}$ & $c=4.58$, $s=0.26$, $\mu=9.42$, $\sigma=3.91$ & 0.95\\ 
$c_{ph}$  & Normal & $\phi(x; \mu, \sigma)$ & $\mu=1.609$, $\sigma=0.008$ & 0.62 \\
$k$ & Skewed Normal & $2\phi(x; \mu, \sigma)\cdot \Phi(-a\cdot x; \mu, \sigma)$  & $\mu=190.39$, $\sigma=36.02$, $a=-3.68$ & 0.94 \\ 
$\bar{\gamma}$ & Generalized Gamma & $\frac{|c|y^{ca-1}\exp(-y^c)}{\Gamma(a)}$  & $y=(x-\mu)/\sigma$, $a=2.53$, $c=8.55\times10^7$ & 0.92 \\
 & &   & $\mu=-2.11\times10^7$, $\sigma=2.11\times10^7$ &  \\
$\alpha$  &  Johnson SU  & $\frac{b}{\sqrt{x^2+1}}\phi\left(a+b\log(x+\sqrt{x^2+1}); \mu, \sigma\right)$ & $a=1.33$, $b=2.00$, $\mu=3.54$, $\sigma=1.43$ & 0.73 \\
\end{tabular} 
\end{ruledtabular}
\end{table}
\endgroup

\newpage

\section{Finite difference formulas}

\subsection{2nd derivatives}

The harmonic force constants are calculated by numerically approximating the second derivative $\Phi_{ij}^{\alpha\beta} = \partial^2 U/\partial u_i^\alpha \partial u_j^\beta, $ where $U$ is the energy of the system and $u_i^\alpha$ and $u_j^\beta$ are displacements of atoms $i$ and $j$ in directions $\alpha$ and $\beta$ in the supercell. Using the shorthand $U(u_i^\alpha\pm h, u_j^\beta \pm h)=U_{i\pm h, j\pm h}$, where $h$ is the magnitude of the atomic displacement, the derivatives are approximated as

\begin{equation}\label{eqn: fd}
\Phi_{ij}^{\alpha\beta}\approx
\begin{cases} 
\frac{1}{4h^2}\bigg(U_{i+h,j+h}+U_{i-h,j-h}-U_{i+h,j-h}-U_{i-h,j+h}\bigg),&\text{if}\ i\neq j\\
\frac{1}{h^2}\bigg( U_{i+h} - 2U_0+U_{i-h} \bigg),  &\text{if}\ i=j.\\
\end{cases}
\end{equation}

where $U_0$ is the ground state energy of the system (i.e., where there are no displacements).

\subsection{3rd derivatives}

Using a similar shorthand as in the previous subsection, the third-order finite difference formulas are provided below. Note that the first formula is symmetric with respect to permutations. For example, the case presented is $u_i^\alpha =u_k^\gamma$ and $u_i^\alpha\neq u_j^\beta$. If instead $u_i^\alpha =u_j^\beta$ and $u_i^\alpha\neq u_k^\gamma$, the perturbed variable would change but the formula would otherwise remain the same. That is, $U_{i+h, j+h,k}$ would be replaced by $U_{i+h, j, k+h}$, and similar substitutions would be made for the other terms.

\begin{equation}\label{eqn: fd3}
\Psi_{ijk}^{\alpha\beta\gamma}\approx
\begin{cases} 
\frac{1}{2h^3}\bigg(U_{i+h,j+h,k} +U_{i-h,j+h,k}+2U_{i,j-h,k}-U_{i-h,j-h,k}-U_{i+h,j-h,k}-2U_{i,j+h,k} \bigg),&\text{if}\ i\neq j, i=k\\
\frac{1}{2h^3}\bigg(-U_{i-2h,j,k} + 2U_{i-h,j,k}-2U_{i+h,j,k}+U_{i+2h,j,k} \bigg),  &\text{if}\ i=j=k\\
\frac{1}{8h^3}\bigg(U_{i+h,j+h,k+h}+U_{i+h,j-h,k-h}+U_{i-h,j-h,k+h}+U_{i-h,j+h,k-h}-U_{i-h,j-h,k-h}-\\
\hspace{0.75cm} U_{i-h,j+h,k+h}-U_{i+h,j-h,k+h}-U_{i+h,j+h,k-h} \bigg),  &\text{if}\ i\neq j\neq k.\\
\end{cases}
\end{equation}

\newpage

\section{Thermal conductivity prediction using forces}

In order to use the BEEF-vdW ensemble, we calculate interatomic force constants using a central finite difference of the energies of structures perturbed from equilibrium. It is more common to calculate force constants using forces, such as in Eqs.`2.58(a) and 2.58(b) from Jain.\cite{ankit_thesis} We show here the lattice thermal conductivity results for silicon with force constants calculated using forces rather than energies. These calculations were performed using the LDA XC functional in GPAW with the computational parameters listed in Sec II C. The lack of significant variation between the predictions indicates that using energies to calculate the force constants does not hurt our prediction accuracy.

\begingroup
\squeezetable
\begin{table}[h]
\caption{\label{tbl: si forces}TA phonon frequency at the $X$ point, [100] LA sound speed, three-phonon phase space, average Gr{\"u}neisen parameter, and thermal conductivity (at $T=300$ K) of silicon with force constants calculated using a finite difference of energies or forces.} % QE is all dfpt
\begin{ruledtabular}
\begin{tabular}{>{\RaggedLeft\arraybackslash}p{2.75cm}|>{\RaggedLeft\arraybackslash}p{2cm}|>{\RaggedLeft\arraybackslash}p{2cm}|>{\RaggedLeft\arraybackslash}p{1.7cm}|>{\RaggedLeft\arraybackslash}p{1.75cm}|>{\RaggedLeft\arraybackslash}p{2.1cm}}
 Force constant calculation & TA frequency at $\mathbf{X}$-pt (THz) & LA [100] sound speed (m/s) & Three-phonon phase space ($\times 10^{-3})$ & Avg. Gr{\"u}neisen parameter & Thermal conductivity (W/m-K)  \\
\hline
Energies &  3.97 & 8388 & 1.23 & 1.16 &  122 \\
Forces & 3.98 & 8315 & 1.23 & 1.15 & 121  \\
\end{tabular}
\end{ruledtabular}
\end{table}
\endgroup

\newpage

\section{Three-phonon phase space}

The three-phonon phase space $P_3$ is a useful harmonic-level property that quantifies, based on the phonon dispersion, the number of three-phonon scattering processes that satisfy energy and momentum conservation. \cite{lindsay2} It is defined as 

\begin{equation}\label{eqn: three phonon 1}
P_3 = \frac{2}{3\Omega} \left(P_3^{(+)}+\frac{1}{2}P_3^{(-)} \right),
\end{equation}

\noindent where 

\begin{equation}\label{eqn: three phonon 2}
P_3^{(\pm)} = \sum_\nu \int d\mathbf{q}D_\nu^{(\pm)}(\mathbf{q})
\end{equation}

\noindent and 

\begin{equation}\label{eqn: three phonon 3}
D_\nu^{(\pm)} (\mathbf{q}) = \sum_{\nu',\nu''}\int  d\mathbf{q}'\delta [\omega_\nu(\mathbf{q})\pm \omega_{\nu'}(\mathbf{q}')-\omega_{\nu''}(\mathbf{q}\pm\mathbf{q}'-\mathbf{G})].
\end{equation}

\noindent In Eqs. (\ref{eqn: three phonon 1})-(\ref{eqn: three phonon 3}), $\mathbf{G}$ is a reciprocal lattice vector, $\Omega$ is a normalization factor, and $D_\nu^{(\pm)}$ is the two-phonon density of states for three-phonon processes involving phonon mode $\nu$ (the $\pm$ distinguishes Type I and Type II three-phonon processes). \cite{lindsay2, shengbte}

\bibliography{bibliography}

\newpage